\documentclass{article}
\usepackage[utf8]{inputenc} 
\usepackage[T1]{fontenc}    
\usepackage{lmodern}
\usepackage{hyperref}
\usepackage{booktabs}       
\usepackage{amsfonts}       
\usepackage{nicefrac}       
\usepackage{microtype}      
\usepackage{lipsum}
\usepackage[autolanguage]{numprint}
\usepackage{amsmath}
\usepackage{amssymb}
\usepackage{mathtools}
\usepackage{bm}
\usepackage{graphicx}
\usepackage{setspace}
\usepackage{authblk}

\usepackage{floatrow}
\newfloatcommand{capbtabbox}{table}[][\FBwidth]

\usepackage{subcaption}
\usepackage{geometry}
\usepackage{soul}
\usepackage{color}
\newcommand{\ml}[1]{\textcolor{black}{#1}}
\usepackage{mdframed}

\begin{document}

\title{Accelerating Uncertainty Quantification of Groundwater Flow Modelling Using a Deep Neural Network Proxy}

\author[1]{Mikkel B. Lykkegaard\thanks{Address for correspondence: Mikkel Bue Lykkegaard, College of Engineering, Mathematics and Physical Sciences, Harrison Building, University of Exeter, EX4 4QF, UK, E-mail: \texttt{m.lykkegaard@exeter.ac.uk}}}
\author[1,2]{Tim J. Dodwell}
\author[1]{David Moxey}

\affil[1]{CEMPS, University of Exeter, UK.}
\affil[2]{The Alan Turing Institute, London, NW1 2DB, UK.}

\maketitle


\begin{abstract}
Quantifying the uncertainty in model parameters and output is a critical component in model-driven decision support systems for groundwater management. This paper presents a novel algorithmic approach which fuses Markov Chain Monte Carlo (MCMC) and Machine Learning methods to accelerate uncertainty quantification for groundwater flow models. We formulate the governing mathematical model as a Bayesian inverse problem, considering model parameters as a random process with an underlying probability distribution. MCMC allows us to sample from this distribution, but it comes with some limitations: it can be prohibitively expensive when dealing with costly likelihood functions, subsequent samples are often highly correlated, and the standard Metropolis-Hastings algorithm suffers from the curse of dimensionality. 
This paper designs a Metropolis-Hastings proposal which exploits a deep neural network (DNN) approximation of \ml{a groundwater flow model}, to significantly accelerate \ml{MCMC sampling}. We modify a delayed acceptance (DA) model hierarchy, whereby proposals are generated by running short subchains using an inexpensive DNN approximation, resulting in a decorrelation of subsequent fine model proposals. Using a simple adaptive error model, we estimate and correct the bias of the DNN approximation with respect to the posterior distribution on-the-fly. The approach is tested on two synthetic examples; a isotropic two-dimensional problem, and an anisotropic three-dimensional problem. The results show that the cost of uncertainty quantification can be reduced by up to \ml{50\%} compared to single-level MCMC, depending on the precomputation cost and accuracy of the employed DNN.
\end{abstract}


\section{Introduction}

Modelling of groundwater flow and transport is an important decision support tool when, for example, estimating the sustainable yield of an aquifer or remediating groundwater pollution. However, the input parameters for mathematical models of groundwater flow (such as subsurface transmissivity and boundary conditions) are often impossible to determine fully or accurately, and are hence subject to various uncertainties. In order to make informed decisions, it is of critical importance to decision makers to obtain robust and unbiased estimates of the total model uncertainty, which in turn is a product of the uncertainty of these input parameters \cite{anderson_applied_2015}. A popular way to achieve this, in relation to groundwater flow or any inverse problem in general, is stochastic or Bayesian modelling \cite{woodbury_full-bayesian_2000, mariethoz_bayesian_2010, de_la_varga_structural_2016}. In this context, a probability distribution, the {\em prior}, is assigned to the input parameters, in accordance with any readily available information. Given some real-world measurements corresponding to the model outputs (e.g. sparse spatial measurements of hydraulic head, Darcy flow or concentration of pollutants), it is possible to reduce the overall uncertainty and obtain a better representation of the model by conditioning the prior distribution on this data. The result is a distribution of the model input parameters given data, which is also referred to as the \emph{posterior}.

\smallskip
Obtaining samples from the posterior distribution directly is not possible for all but the simplest of problems. A popular approach for generating samples is the  Metropolis–Hastings type {\em Markov Chain Monte Carlo} (MCMC) method \cite{robert_monte_2010}. Samples are generated by a sequential process. First, given a current sample, a new proposal for the input parameters is made using a so-called proposal distribution. Evaluating the model with this new set of parameters, a {\em likelihood} is computed - a measure of misfit between the model outputs and the data. The likelihood\ml{s} of the proposed and current samples are then compared. Based on this comparison, the proposal is either accepted or rejected, and the whole process is repeated, generating a Markov chain of probabilistically feasible input parameters. The key point is that the distribution of samples in the chain converges to the {\em posterior} -- the distribution of input parameters given the data \cite{robert_monte_2010}. This relatively simple algorithm can lead to extremely expensive Bayesian computations for three key reasons. First, each step of the chain requires the evaluation of (often) an expensive mathematical model. Second, the sequential nature of the algorithm means subsequent samples are often highly correlated -- even repeated if a step is rejected. Therefore the chains must often be very long to obtain good statistics on the distribution of outputs of the model. Third, without special care, the approach does not generally scale well to large numbers of uncertain input parameters; the so-called curse of dimensionality. Addressing these scientific challenges is at the heart of modern research in MCMC algorithms. As with this paper there is a particular focus on developing novel and innovative proposal distributions, which seek to de-correlate adjacent samples and limit the computational burden of evaluating expensive models. 

\smallskip
Broadly in the literature, simple Darcy type models and other variants of the diffusion equation have long been a popular toy example problems for demonstrating MCMC methodologies in the applied mathematics community (see e.g. \cite{higdon_markov_2003, dodwell_hierarchical_2015, detommaso_continuous_2018}). There appears to be much less interest in MCMC in the applied groundwater modelling community. This may be because of the computational cost of running MCMC on highly parametrised, expensive models, or the lack of an easy-to-use MCMC software framework, akin to the parameter estimation toolbox PEST \cite{doherty_calibration_2015}. 

\smallskip
An exciting approach to significantly reduce the computational cost has been proposed in multi-level, multi-fidelity and Delayed Acceptance (DA) MCMC methods. In each case, to alleviate computational cost, a hierarchy of models is established, consisting of a fine model and (possibly multiple) coarse, computationally cheap approximations. Typically, the coarser models are finite element solutions of the PDE on a mesh with a coarser resolution, but as we show in this paper, can be taken to be any general approximation similar to the multi-fidelity philosophy \cite{Peh18}. Independent of the approach, the central idea is the same: to obtain significant efficiency gains by exploiting approximate coarse models to generate `good' proposals cheaply, using additional accept/reject steps to filter out highly unlikely proposals before evaluating the fine, expensive model. Previous studies of two-stage approaches include \cite{efendiev_efficient_2005} who modelled multi-phase flow with coarse level proposals evaluated by a coarse-mesh single-phase flow model (an idea that was developed further in \cite{mondal_bayesian_2010}), \cite{dostert_efficient_2009} and  \cite{laloy_efficient_2013}. We note that the latter of which, instead of simply using a coarser discretisation, implemented a data-driven polynomial chaos expansion as a surrogate model. We intend to demonstrate how the development of novel techniques in MCMC and machine learning can be combined to help realise the potential of MCMC in this field.

\smallskip
In this work, we propose a combination of multiple cutting-edge MCMC techniques to allow for efficient inversion and uncertainty quantification of groundwater flow. We propose an improved delayed acceptance (DA) MCMC algorithm, adapted from the approach proposed by \cite{christen_markov_2005}. In our case, similarly to multi-level MCMC \cite{dodwell_hierarchical_2015}, proposals are generated by computing a subchain using a Deep Neural Network (DNN) as an approximate model -- leading to cheaply computed, decorrelated proposals passed on to the fine model. For our first example, the subchain is driven by the preconditioned Crank-Nicolson (pCN) proposal distribution \cite{cotter_mcmc_2013} to ensure the proposed Metropolis-Hastings algorithm is robust with respect to the dimension of the uncertain parameter space. For our second example, proposals for the subchains are generated using the Adaptive Metropolis (AM) proposal \cite{haario_adaptive_2001}, since the posterior distribution in this case is highly non-spherical and multiple parameters are correlated. Finally, we propose a enhanced error model, in which the DNN is trained by sampling the prior distribution, yet the bias of the approximation is adaptively estimated and corrected on-the-fly by testing the approximations against the full model in an adaptive delayed acceptance setting \cite{cui_posteriori_2018}. 

\section{Preliminaries}
\label{sec:preliminaries}
In this section we briefly introduce the forward model, defining the governing equations underpinning groundwater flow and their corresponding weak form, enabling us to solve the equations using FEM methods. We then formulate our model as a Bayesian inverse problem with random input parameters, effectively resulting in a stochastic model, which can be accurately characterised by sampling from the posterior distribution of parameters using MCMC. The simple Metropolis-Hastings MCMC algorithm is then introduced and extended with the preconditioned Crank-Nicolson (pCN) and Adaptive Metropolis (AM) transition kernels.

\subsection{Governing equations for groundwater flow}
Consider steady groundwater flow in a confined, inhomogeneous aquifer which occupies the domain $\Omega$ with boundary $\Gamma$. Assuming that water is incompressible, the governing equations for groundwater flow can be written as the scalar elliptic partial differential equation:
\begin{equation} \label{eq:gwflow}
    - \nabla \cdot (- T({\bf x}) \nabla h({\bf x})) = g({\bf x}) \quad \mbox{for all} \quad {\bf x} \in \Omega
\end{equation}
subject to boundary conditions on $\Gamma = \Gamma_N \cup \Gamma_D$ defined by the constraint equations
\begin{equation}
                                h({\bf x})  = h_D({\bf x}) \quad \text{on} \; \Gamma_D \quad\mbox{and}\quad (- T({\bf x}) \nabla h({\bf x})) \cdot \bm{n} = q_N({\bf x}) \quad \text{on} \; \Gamma_N.
\end{equation}
Here $T({\bf x})$ is the heterogeneous, depth-integrated transmissivity, $h ({\bf x})$ is hydraulic head, $h_D ({\bf x})$ is fixed hydraulic head at boundaries with Dirichlet constraints, $g({\bf x})$ is fluid sources and sinks, $q({\bf x})$ is Darcy velocity, $q_N ({\bf x})$ is Darcy velocity across boundaries with Neumann constraints and $\Gamma_D\subset\partial\Omega$ and $\Gamma_N\subset\partial\Omega$ define the boundaries comprising of Dirichlet and Neumann conditions, respectively. Following standard FEM practice (see e.g. \cite{diersch_feflow:_2014}), eq.~\eqref{eq:gwflow} is converted into weak form by multiplying by an appropriate test function $w\in H^1(\Omega)$ and integrating by parts, so that
\begin{equation}\label{eqn:weakform}
 \int_{\Omega} \nabla w \cdot \left(T({\bf x}) \nabla h \right) \: d{\bf x} + \int_{\Gamma_N} w \: q_N({\bf x}) \: ds = \int_{\Omega} w \: g({\bf x}) \: d{\bf x}, \quad \forall w \in H^1(\Omega),
\end{equation}
where $H^1(\Omega)$ is the Hilbert space of weakly differentiable functions on $\Omega$. To approximate the hydraulic head solution $h(\bm{x})$, a finite element space $V_\tau \subset H^1(\Omega)$  on a finite element mesh $\mathcal Q_\tau(\Omega)$. This is defined by a basis of piecewise linear Lagrange polynomials $\{\phi_i({\bf x})\}_{i=1}^M$, associated with each of the $M$ finite element nodes.  As a result \eqref{eqn:weakform} can be rewritten as a system of sparse linear equations
\begin{align}
    {\bf A} {\bf h} = {\bf b} \quad \mbox{where} \quad A_{ij} &= \int_\Omega \nabla \phi_i \cdot T({\bf x}) \nabla \phi_j({\bf x})\;d{\bf x} \quad \mbox{and} \\
                                                       b_i &= \int_{\Omega} \phi_i({\bf x}) \: g({\bf x}) \: d{\bf x} - \int_{\Gamma_N} \phi_i({\bf x}) q_N({\bf x}) \: ds,
\end{align}
where ${\bf A} \in \mathbb R^{M \times M}$ and ${\bf b} \in \mathbb R^M$ are the global stiffness matrix and load vector, respectively. The vector ${\bf h} := [h_1, h_2, \ldots, h_M] \in \mathbb R^M$ is the solution vector of hydraulic head at each node within the finite element mesh so that $h({\bf x}) = \sum_{i=1}^M h_i \phi_i({\bf x})$. In our numerical experiments, these equations are solved using the open source general-purpose FEM framework \texttt{FEniCS} \cite{langtangen_solving_2017}. While there are well-established groundwater simulation software packages available, such as MODFLOW \cite{harbaugh_modflow-2005_2005} and FEFLOW \cite{diersch_feflow:_2014}, FEniCS was chosen because of its flexibility and ease of integration with other software and analysis codes. 

\subsection{Aquifer transmissivity}

The aquifer transmissivity $T({\bf x})$ is not known everywhere on the domain, therefore a typical approach is to model it as a log-Gaussian random field. There exists extensive literature on modelling groundwater flow transmissivity using  log-Gaussian random fields (see e.g. \cite{freeze_stochastic-conceptual_1975, kitterrod_simulation_1997, laloy_efficient_2013}). Whilst this may not always prove a good model, particularly in cases with highly correlated extreme values and/or preferential flow paths \cite{gomez-hernandez_be_1998, kerrou_issues_2008} as seen when considering faults and other discontinuities \cite{russo_statistical_1992, hoeksema_analysis_1985}. However, the log-Gaussian distribution remains relevant for modelling transmissivity in a range of aquifers \cite{dostert_coarse-gradient_2006, marzouk_dimensionality_2009, laloy_efficient_2013}. 

Our starting point is a covariance operator with kernel $C({\bf x},{\bf y})$, which defines the correlation structure of the uncertain transmissivity field. 
For our numerical experiments, we consider the ARD (Automatic Relevance Determination) squared exponential kernel, a generalisation of the `classic' squared exponential kernel, which allows for handling directional anisotropy:
\begin{equation}
C({\bf x}, {\bf y}) = \exp \left( - \frac{1}{2} \sum_{j=1}^{d} \left(\frac{x_j - y_j}{l_j} \right)^2 \right),
\end{equation}
where $d$ is the spatial dimensionality of the problem and $\bm{l} \in \mathbb R^{d}$ is a vector of lengths scales corresponding to each spatial dimension. We emphasise that the covariance kernel is a \emph{modelling choice}, and that different options are available, such as the Matern kernel which offers additional control over the smoothness of the field. 

In our work, transmissivity was modeled as a discrete log-Gaussian random field expanded in an orthogonal eigenbasis with $k$ Karhunen-Loève (KL) eigenmodes. To achieve this we construct a covariance matrix ${\bf C} \in \mathbb R^{M\times M}$, where entries are given by $C_{ij} = C({\bf x}_i, {\bf x}_j)$ for each pair of nodal coordinates within the finite element mesh $i,j = 1, \ldots M$. Once constructed the largest $k$ eigenvalues $\{\lambda_i\}_{i=1}^k$ and associated eigenvectors $\{\bm \psi_i\}_{i=1}^k$ of ${\bf C}$ can be computed. The transmissivity at the nodes ${\bf t} := [t_1, t_2, \ldots, t_M]$, is given by
\begin{equation}
    \log \bm{t} = {\bm \mu} + \sigma \bm{\Psi} \bm{\Lambda}^{\frac{1}{2}} \bm{\theta}, \quad \mbox{where} \quad {\bf \Lambda} =
     \begin{bmatrix}
    \lambda_1 & 0 & \dots & 0 \\
    0 & \lambda_2 & \dots & 0 \\
    \vdots & \vdots & \ddots & \vdots \\
    0 & 0 & \dots & \lambda_k
  \end{bmatrix}\quad \mbox{and} \quad \bm{\Psi} = [{\bm \psi}_1, {\bm \psi}_2, \ldots, {\bm \psi}_k],
\end{equation}
vector $\bm{\mu}$ defines the log of the mean transmissivitity field, $\sigma$ a scalar parametrising the variance and $\bm{\theta}$ a vector of Gaussian random variables such that $\bm{\theta} \sim \mathcal{N}(0,\mathbb{I}_k)$ as in \cite{scarth_random_2019}. The random field can be interpolated from nodal values across $\Omega$, using the shape functions $\{\phi_i({\bf x})\}_{i=1}^M$ so that $T({\bf x}) = \sum_{i=1}^M t_i \phi_i({\bf x}).$ 

\begin{figure}[!h]
    \centering
    \includegraphics[width=0.75\textwidth]{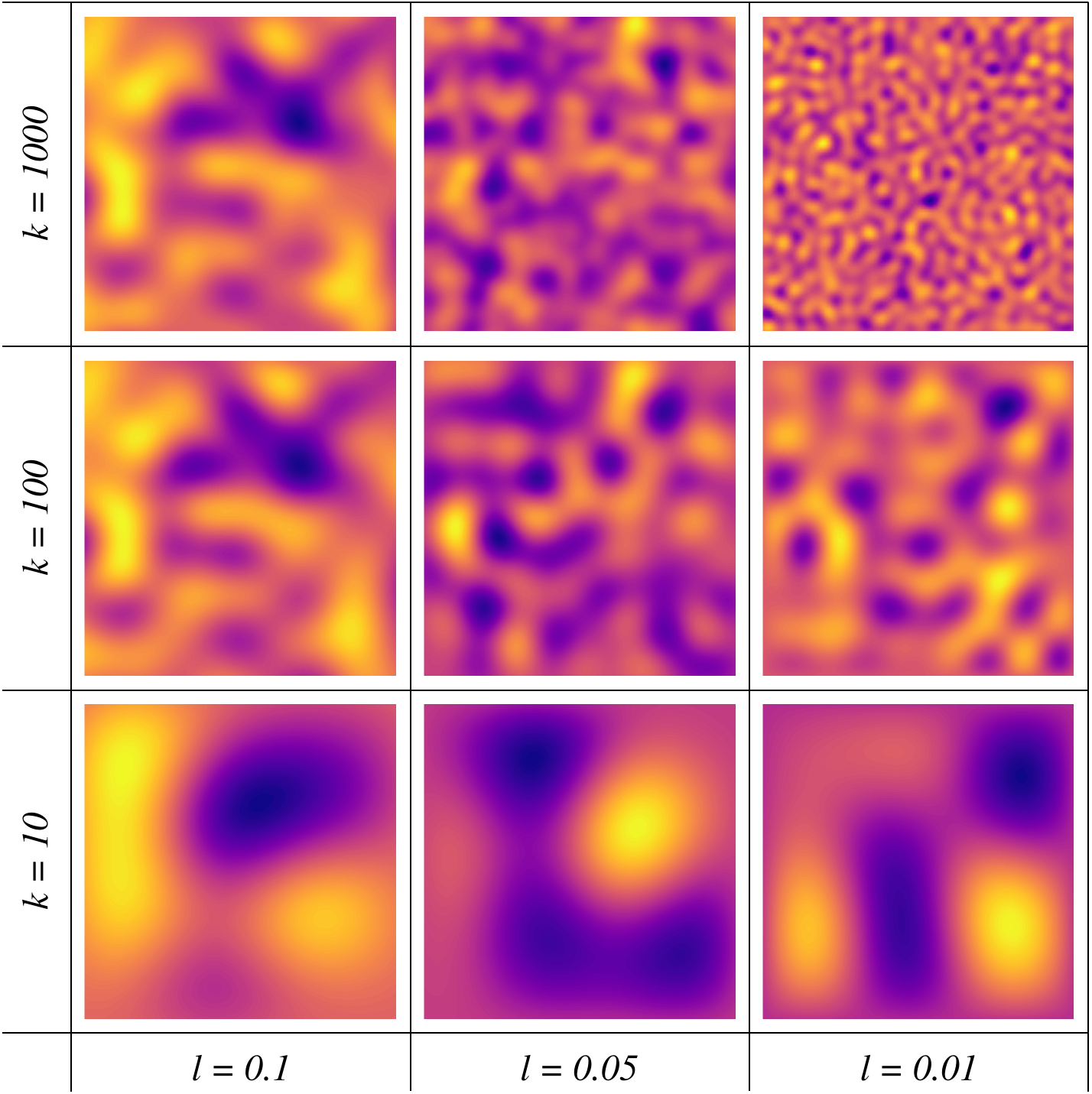}
    \caption{A selection of Gaussian random process realisations \ml{for $\bm{x} \in [0,1]^2$}, with a square exponential kernel using different covariance length scales $l$ and number of KL eigenmodes $k$. All displayed realisations were generated using the same appropriately truncated random vector $\bm{\xi}$ with identical eigenvectors for each $l$.}
    \label{fig:random_fields}
\end{figure}

Truncating the KL eigenmodes at the $k$th mode \ml{limits the amount of small scale features that can be represented. This, along with} interpolating the field, has a smoothing effect on the recovered transmissivity fields, which may or may not be desirable, depending on the application. Figure \ref{fig:random_fields} shows some examples of realisations of Gaussian random fields with a square exponential kernel, which illustrates the effect of the covariance length scale $l$ and the number of admitted KL eigenmodes $k$. \ml{For relatively large length scales $l$, there is a limit to $k$, above which adding higher frequency eigenvalues does not provide any additional information. In this context, the proportion of signal energy encompassed by the truncation can be understood as the ratio between the sum of truncated eigenvalues and the sum of all eigenvalues: $\sum_{i=1}^{k} \lambda_i / \sum_{j=1}^{M} \lambda_j$.}

\subsection{The Bayesian inverse problem}
To setup the Bayesian inverse problem and thereby quantify the uncertainty in the transmissivity field $T({\bf x})$, the starting point is \ml{to} define a statistical model which describes distribution of the mismatch between observations and model predictions.  The observations are expressed in a single vector ${\bf d}_{obs} \in \mathbb R^m$ and for a given set of model input parameters ${\bm \theta}$, the model's prediction of the data is defined by the \textit{forward map}, $\mathcal F({\bm \theta}): \mathbb R^k \rightarrow \mathbb R^m$. The statistical model assumes the connection between model and observations through the relationship
\begin{equation} \label{eq:inverse_problem}
\bm{d}_{\text{obs}}= \mathcal{F}(\bm{\theta}) + \bm{\epsilon}
\end{equation}
where we take $\bm{\epsilon} \sim N(0,{\bm\Sigma}_{\epsilon})$ which represents the uncertainty \ml{of} the connection between model and data, capturing both model mis-specification and measurement noise as sources of this uncertainty.

\smallskip
The backbone of a Bayesian approach is Bayes' theorem, which allows for computing \textit{posterior} beliefs of model parameters using both \textit{prior} beliefs and \textit{observations}. Bayes' theorem states that the posterior probability of a parameter realisation $\bm{\theta}$ given data $\bm{d}_{\text{obs}}$ can be computed as
\begin{equation} \label{eq:bayes}
    \pi(\bm{\theta} | \bm{d}_{\text{obs}}) = \frac{\pi_0(\bm{\theta}) \mathcal{L}(\bm{d}_{\text{obs}} | \bm{\theta})}{\pi(\bm{d}_{\text{obs}})}
\end{equation}
where $\pi(\bm{\theta} | \bm{d}_{\text{obs}})$ is referred to as the \textit{posterior distribution}, $\mathcal{L}(\bm{d}_{\text{obs}} | \bm{\theta})$ is called the \textit{likelihood}, $\pi_0(\bm{\theta})$ the \textit{prior distribution} and
\begin{equation}\label{eqn:evidence}
    \pi(\bm{d}_{\text{obs}}) = \int_{\Theta} \pi(\bm{d}_{\text{obs}}| \bm{\theta}) d \bm{\theta} = \int_{\Theta} \pi_0(\bm{\theta}) \mathcal{L}(\bm{d}_{\text{obs}} | \bm{\theta}) d \bm{\theta}
\end{equation}
is a normalising constant, sometimes referred to as the \textit{evidence}. In most cases this integral does not have a closed-form solution and is infeasible to estimate numerically in most real-world applications, particularly when the dimension of the unknown parameter space is large and the evaluation of the model (required to compute $\mathcal L({\bf d}_{obs}|{\bm \theta})$) is computationally expensive. A family of methods called Markov Chain Monte Carlo (MCMC) are often employed to approximate the solution \cite{gelman_bayesian_2004}. Importantly MCMC, whilst computationally expensive, allows indirect sampling from the posterior distribution and avoids the explicit need to estimate \eqref{eqn:evidence}. Moreover, it can be designed to be independent of the dimension of the parameter space and has no embedded unquantifiable bias. In this paper we consider a subclass of MCMC methods called the Metropolis-Hastings \cite{metropolis_equation_1953,hastings_monte_1970,robert_monte_2010} algorithm, which is described in Algorithm 1. The algorithm generates a Markov chain $\{{\bm \theta}^{(n)}\}_{n \in \mathbb N}$ with a distribution converging to $\pi({\bm d}_{obs}|{\bm \theta})$. It is difficult (often impossible) to sample directly from the posterior, hence at each step, at position ${\bm \theta}^{(i)}$ in the chain, a proposal is made ${\bm \theta}'$ from a simpler known (proposal) distribution $q(\bm{\theta}'|\bm{\theta}^{(i)})$. An accept/reject step then determines whether the proposal comes from (probabilistically) the posterior distribution or not. This accept/reject step is a achieved by essentially computing the ratio of the densities of the current state to the proposal. To do this we exploit Bayes's Theorem. The key observation in MCMC is that the normalising constant $\pi(\bm{d}_{\text{obs}})$ is independent of $\bm{\theta}$, and so
\begin{equation}
    \pi(\bm{\theta} | \bm{d}_{\text{obs}}) \propto \pi_0(\bm{\theta}) \mathcal{L}(\bm{d}_{\text{obs}} | \bm{\theta}).
\end{equation}
Therefore when comparing the ratio of the densities, the normalizing constant (since independent of ${\bm \theta}$) cancels.
\begin{center}
    \fbox{\parbox{\textwidth}{
            \textbf{Algorithm 1: Metropolis-Hastings Algorithm}
            \begin{enumerate}
                \item Given a parameter realisation $\bm{\theta}_i$ and a transition kernel $q(\bm{\theta}' | \bm{\theta}_i)$, generate a proposal $\bm{\theta}'$.
                \item Compute the likelihood ratio between the proposal and the previous realisation: 
                \begin{equation*}
                \alpha = \text{min} \left\{ 1, \frac{\pi_0(\bm{\theta}') \mathcal{L}(\bm{d}_{\text{obs}} | \bm{\theta}')}{\pi_0(\bm{\theta}^{(i)}) \mathcal{L}(\bm{d}_{\text{obs}} | \bm{\theta}^{(i)})} \frac{q(\bm{\theta}^{(i)}|\bm{\theta}')}{q(\bm{\theta}'|\bm{\theta}^{(i)})} \right\}
                \end{equation*} 
                \item If $u \sim U(0,1) > \alpha$ then set $\bm{\theta}^{(i+1)} = \bm{\theta}^{(i)}$, otherwise, set $\bm{\theta}^{(i+1)} = \bm{\theta}'$.
            \end{enumerate}
    }}
\end{center}

In our model problem, the prior density of the parameters $\pi_0(\bm{\theta})$ represents the available \textit{a priori} knowledge about the transmissivity of the aquifer. From our statistical model \eqref{eq:inverse_problem} we see that our ${\bm d}_{obs} - \mathcal F({\bm \theta}) \sim \mathcal N(0,{\bm \Sigma}_\epsilon)$, hence 
\begin{equation} \label{eq:likelihood}
\mathcal{L}(\bm{d}_{\text{obs}} | \bm{\theta}) = \exp \left (-\frac{1}{2} (\mathcal{F}(\bm{\theta}) - \bm{d}_{\text{obs}})^\intercal \bm{\Sigma}_e^{-1} (\mathcal{F}(\bm{\theta}) - \bm{d}_{\text{obs}}) \right).
\end{equation}
Importantly we note that for each step of the Metropolis-Hastings algorithms we are required to compute $\mathcal L({\bm d}_{obs}|{\bm \theta}')$. This requires the evaluation of the forward mapping $\mathcal F({\bm \theta}')$ which can be computationally expensive. Moreover, due to the sequential nature of MCMC-based approaches, consecutive samples are correlated and hence many samples are required to obtain good statistics on the outputs.

The proposal distribution $q({\bm \theta}'|\bm \theta^{(n)})$ is the key element which drives the Metropolis-Hastings algorithm and control the effectiveness of the algorithm. A common choice is a simple random walk, for which $q_{\text{RW}}(\bm{\theta}'|\bm{\theta}^{(i)}) = \mathcal{N}(\bm{\theta}^{(i)},\bm{\Sigma})$, yet as shown in \cite{katafygiotis_geometric_2008} 
, the basic random walk does not lead to a convergence that is independent of the input dimension $m$. Better choices would be the \textit{preconditioned Crank-Nicolson} proposal (pCN, \cite{cotter_mcmc_2013}), which has dimension independent acceptance probability, or the \textit{Adaptive Metropolis} algorithm (AM, \cite{haario_adaptive_2001}), which adaptively aligns the proposal distribution to the posterior during sampling. Moreover, unlike the Metropolis-Adjusted Langevin Algorithm (MALA), No-U-Turn Sampler (NUTS) and Hamiltonian Monte Carlo, none of these proposals rely on gradient information, which can be infeasible to compute for expensive forward models. 

To generate a proposal using the pCN transition kernel, one computes
\begin{equation}
    \bm{\theta}' = \sqrt{1-\beta^2}\,\bm{\theta}^{(i)} + \beta \bm{\xi}
\end{equation}
where $\bm{\xi}$ is a random sample from the prior distribution, $\bm{\xi} \sim \mathcal{N}(0,\bm{\Sigma})$. This expression corresponds to the transition kernel $q_{\text{pCN}}(\bm{\theta}'|\bm{\theta}^{(i)}) = \mathcal{N}(\sqrt{1-\beta^2}\bm{\theta}^{(i)}, \beta \bm{\Sigma})$. Moreover, for the pCN transition kernel, the acceptance probability simplifies to 
\begin{equation}
    \alpha = \text{min} \left\{ 1, \frac{\mathcal{L}(\bm{d}_{\text{obs}} | \bm{\theta}')}{\mathcal{L}(\bm{d}_{\text{obs}} | \bm{\theta}^{(i)})} \right\}
\quad \mbox{following the identity} \quad
  \frac{p_0(\bm{\theta}^{(i)})}{p_0(\bm{\theta}')} = \frac{q_{\text{pCN}}(\bm{\theta}^{(i)}|\bm{\theta}')}{q_{\text{pCN}}(\bm{\theta}'|\bm{\theta}^{(i)})}
\end{equation}
as given in \cite{dodwell_hierarchical_2015}. Additional details of derivation of the pCN proposal are are provided in Appendix~\ref{app:pCN}.

Similarly, to generate a proposal using the AM transition kernel, we draw a random sample 
\begin{equation}
\bm{\theta}' \sim \mathcal{N}(\bm{\theta}^{(i)}, \bm{\Sigma}^{(i)})
\end{equation}
where $\bm{\Sigma}^{(i)}$ is an iteratively updated covariance structure
\begin{equation*}
\bm{\Sigma}^{(i)} = 
\begin{cases}
\bm{\Sigma}^{(0)},& \text{if } i \le i_0,\\
s_d \, \text{Cov}(\bm{\theta}^{(0)}, \, \bm{\theta}^{(1)} \, ... \, \bm{\theta}^{(i)}) + s_d \, \gamma \, \mathbb{I}_d, & \text{otherwise}.
\end{cases}
\end{equation*}
Hence, proposals are drawn from a distribution with an initial covariance $\bm{\Sigma}^{(0)}$ for a given period $i_0$, after which adaptivity is 'switched on', and used for the remaining samples. The adaptive covariance $\bm{\Sigma}^{(i)} = s_d \, \text{Cov}(\bm{\theta}^{(0)}, \, \bm{\theta}^{(1)} \, ... \, \bm{\theta}^{(i)}) + s_d \, \gamma \, \mathbb{I}_d$ can be constructed iteratively during sampling using the following recursive formula:
\begin{equation}
    \bm{\Sigma}^{(i+1)} = \frac{i-1}{i} \bm{\Sigma}^{(i)} + \frac{s_d}{i}(i \bar{\bm{\theta}}^{(i-1)} \bar{\bm{\theta}}^{(i-1)\intercal} - (i+1) \bar{\bm{\theta}}^{(i)} \bar{\bm{\theta}}^{(i)\intercal} + \bm{\theta}^{(i)} \bm{\theta}^{(i)\intercal} + \gamma \mathbb{I}_d)
\end{equation}
where $\bar{\cdot}$ is the arithmetic mean, $s_d = 2.4^2/d$ is a scaling parameter, $d$ is the dimension of the proposal distribution and $\gamma$ is a parameter which prevents $\bm{\Sigma}_{i}$ from becoming singular \cite{haario_adaptive_2001}. This, on the other hand, corresponds to the transition kernel $q_{\text{AM}}(\bm{\theta}'|\bm{\theta}^{(0)}, \bm{\theta}^{(1)} \: ... \: \bm{\theta}^{(i)}) = \mathcal{N}(\bm{\theta}^{(i)}, \bm{\Sigma}^{(i)})$, which is not guaranteed to be ergodic, since it will depend on the history of the chain. However, the Diminishing Adaptation condition \cite{roberts_examples_2009} holds, as adaptation will naturally decrease as sampling progresses.

\subsection{Deep Neural Network}
The approximate/surrogate model in our experiments is a feed-forward deep neural network (DNN), a type of artificial neural network with multiple hidden layers, as implemented in the open-source neural-network library \texttt{Keras} \cite{chollet2015keras} utilizing the \texttt{Theano} backend \cite{2016arXiv160502688short}.

\ml{Artificial neural networks have previously been successfully applied as fast model proxies in inverse geophysics problems. Examples include \cite{hansen_efcient_2017}, who used a neural network with two hidden layers for Monte Carlo sampling in the context of a crosshole traveltime inversion, and \cite{moghadas_soil_2020} who used a neural network with a single hidden layer and a Differential Evolution Adaptive Metropolis sampler for electromagnetic inversion.}

\begin{figure}[htbp]
    \centering
    \includegraphics[width=0.4\textwidth]{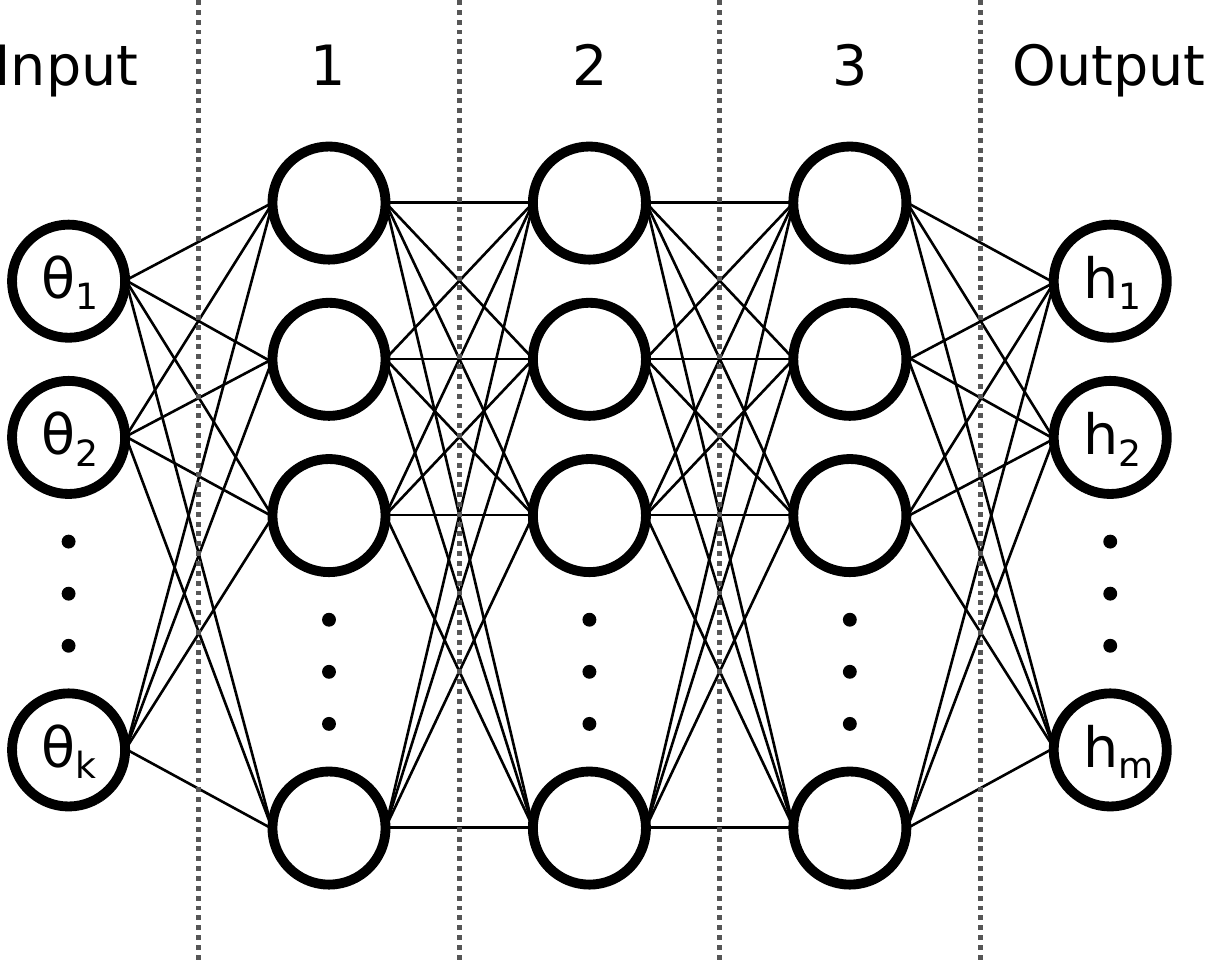}
    \caption{Graph showing the structure of a feedforward DNN.}
    \label{fig:mlp}
\end{figure}

\smallskip
The DNN approximates the forward map, accepting a vector of KL coefficients ${\bm \theta} \in \mathbb R^k$, and returning an approximation of the vector of approximate model output $\mathcal{\hat{F}}(\bm{\theta}) \in \mathbb R^m$ -- in this paper a vector of hydraulic heads at given sampling points, i.e. $\mathcal{\hat{F}}(\bm{\theta}) : \mathbb R^k \mapsto \mathbb R^m $. Figure~\ref{fig:mlp} shows the graph of one particular DNN employed in our experiments. 

Each edge in Figure~\ref{fig:mlp} is equipped with a weight $w^l_{i,j}$ where $l$ is index of the layer that the weight feeds into, $i$ is the index of nodes in the same layer and $j$ is the index of nodes in the previous layer. These weights can be arranged in $n \times m$ matrices $\bm{W}_l$ for each layer $l$. Similarly, each node is equipped with a bias $b^l_i$ where $l$ is index of its layer and $i$ is the index of node, and these biases can be arranged in vectors $\bm{b}_l$. Data is propagated through the network such that the output $\bm{y}_l$ of a layer $l$ with activation function $\mathcal{A}_l(\cdot)$ is
\begin{equation}
\bm{y}_l = \mathcal{A}_l \left( \bm{b}_l + \bm{W}_l \; \bm{y}_{l-1} \right).
\end{equation}
Activation functions $A(\cdot)$ are applied element-wise on their input vectors $\bm{x}$ so that 
\begin{equation*}
    \mathcal{A}(\bm{x}) = (A(x_1), \: A(x_2) \: \dots \: A(x_n))^\intercal
\end{equation*}
Many different activation functions are available for artificial neural networks, and we here give a short description of the ones employed in our experiments: the \textit{sigmoid} and the \textit{rectified linear unit} (`ReLU'). The transfer function of the nodes in the first layer of each DNN was of the type \textit{sigmoid}:
\begin{equation}
    S(x) = \frac{1}{1 + e^{-x}}
\end{equation}
squashing the input vector into the interval $(0, 1)$, effectively resulting in a strictly positive output from the first hidden layer. The remaining hidden layers consisted of nodes with the \textit{de facto} standard hidden layer activation function for deep neural networks, the \textit{rectified linear unit} (`ReLU'):
\begin{equation*}
    R(x) = 
    \begin{cases}
    x,& \text{if } x > 0,\\
    0,              & \text{otherwise}.
    \end{cases}
\end{equation*}

To fit an artificial neural network to a given set of data, the network is initially compiled using random weights and biases and then trained using a dataset of known inputs and their corresponding outputs. The weights and biases are updated iteratively during training by way of an appropriate optimisation algorithm and a loss function, and if appropriately set up, will converge towards a set of optimal values, allowing the DNN to predict the response of the forward model to some level of accuracy \cite{hastie_elements_2009}. Our particular DNNs were trained using the mean squared error (MSE) loss function
\begin{equation*}
    \text{MSE} = \frac{1}{m} \sum_{i = 1}^{m} (h_i - \hat{h}_i)^2
\end{equation*}
for $m$ output variables, and the RMSprop optimiser, a stochastic, gradient based and adaptive algorithm, suggested by \cite{rmsprop} and widely used for training DNNs.

\section{Adaptive Delayed Acceptance Proposal using a Deep Neural Network}
\label{sec:contribution}

In this section we describe a modified adaptive delayed acceptance proposal for MCMC, using ideas from multi-level MCMC \cite{dodwell_hierarchical_2015}. The general approach generates proposals by running Markov subchains driven by an approximate model. In our case this approximation is constructed from a DNN of the forward map $\mathcal F({\bm \theta})$ trained from offline samples of the prior distribution. Finally, we show how the approximate map can be corrected online, by adaptively learning a simple multi-variant Gaussian correction to the outputs of the neural network.

\subsection{Modified Delayed Acceptance MCMC}\label{sec:DA}
Delayed Acceptance (DA) \cite{christen_markov_2005} is a technique that exploits a model hierarchy consisting of an expensive fine model and relatively inexpensive coarse approximation. The idea is simple: a proposal is first evaluated (pre-screened) by an approximate model and immediately discarded if it is rejected. Only if accepted, it is subjected to a second accept/reject step using the fine model. In this context, the likelihood of observations given a parameter set is henceforth denoted $\mathcal{\hat{L}}(\bm{d}_{\text{obs}} | \bm{\theta})$ when evaluated on the approximate model and remains $\mathcal{L}(\bm{d}_{\text{obs}} | \bm{\theta})$ when evaluated on the fine model. This simple screening mechanism cheaply filters out poor proposals, wasting minimal time evaluating unlikely proposals on the expensive, fine model. \ml{Crucially, the coarse model need not evaluate every parameter, only a subset. The remaining fine parameters can then be sampled prior to the second accept/reject step. We denote the full parameter set $\bm{\theta}$, the coarse parameters $\bm{\hat{\theta}}$ and the fine parameters $\bm{\tilde{\theta}}$. so that $\bm{\theta} = [\bm{\hat{\theta}}, \bm{\tilde{\theta}}]$}. 

\begin{figure}[htbp]
    \vspace{0.5cm}
    \centering
    \includegraphics[width=0.5\textwidth]{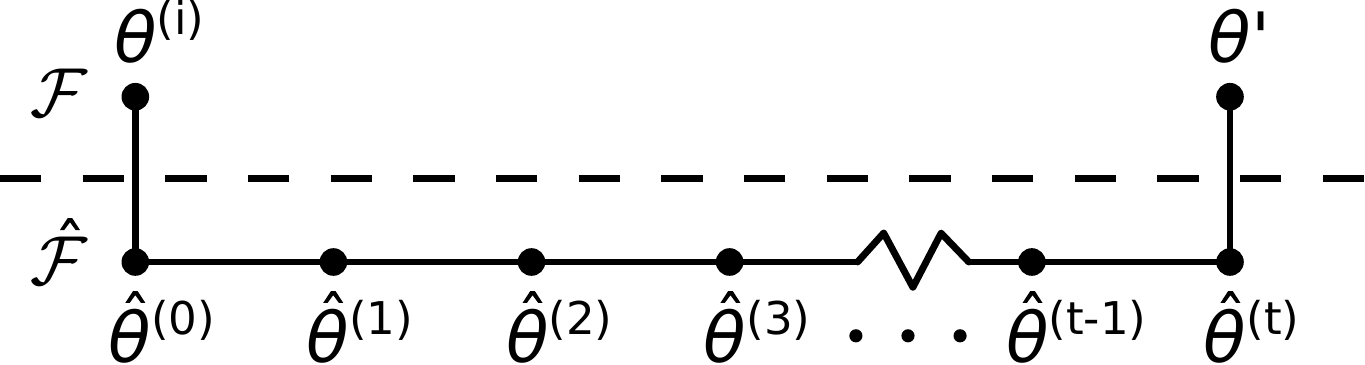}
    \caption{Illustration of the principle used to offset fine level samples to reduce autocorrelation. The fine model $\mathcal{F}$ is only evaluated using the full set of proposed parameters $\bm{\theta}'$ after a prescribed number $t$ \ml{(the \textit{subchain length})} of approximation parameter sets $\{ \bm{\hat{\theta}}^{(1)}, \bm{\hat{\theta}}^{(2)}, \dots, \bm{\hat{\theta}}^{(t)} \}$ have been evaluated on the approximate model $\mathcal{\hat{F}}$ and accepted into the \ml{coarse} chain.}
    \label{fig:da_delay}
\end{figure}

\smallskip
In this paper we extend this approach by not evaluating \emph{every} accepted approximation proposal with the fine model. Instead, a proposal for the fine model is generated by running an approximate subchain until $t$ approximate proposals have been accepted and only then evaluate using the fine model. \ml{We define the required number of accepted proposals in the approximate subchains as the \textit{offset length}}. This modified Delayed Acceptance MCMC algorithm is described in Algorithm 2 and an illustration of the process is given in Figure~\ref{fig:da_delay}.

This way, the autocorrelation of the fine chain is reduced, since proposals are `more independent'. This approach is strongly related to a two-level version of multi-level MCMC. Since the fine model likelihood ratio is corrected by the inverse of the approximate likelihood ratio in step 6 of Algorithm 2, detailed balance is satisfied, the resulting Markov Chain is guaranteed to come from the true posterior and there is no loss of accuracy, even if the approximate model is severely wrong \cite{christen_markov_2005}. To demonstrate that this approach does indeed decrease the autocorrelation in our fine chain MCMC samples, we compute the Effective Sample Size $N_{eff}$ of each MCMC simulation according to the procedures described in \cite{wolff_monte_2007}.

\medskip
\begin{center}
    \fbox{\parbox{\textwidth}{
            \textbf{Algorithm 2: Modified Delayed Acceptance MCMC}
            \begin{enumerate}
                \item Given a realisation of the approximation parameters $\bm{\hat{\theta}}^{(j)}$ and the transition kernel $q(\bm{\hat{\theta}}' | \bm{\hat{\theta}}^{(j)})$, generate a proposal for the approximation $\bm{\hat{\theta}}'$.
                \item Compute the likelihood ratio on the approximate model between the proposal and the previous realisation: 
                \begin{align*}
                \alpha_1 &= \text{min} \left\{ 1, \frac{\pi_0(\bm{\hat{\theta}}') \mathcal{\hat{L}}(\bm{d}_{\text{obs}} | \bm{\hat{\theta}}')}{\pi_0(\bm{\hat{\theta}}^{(j)}) \mathcal{\hat{L}}(\bm{d}_{\text{obs}} | \bm{\hat{\theta}}^{(j)})} \right\} & \text{(\textit{AM})} \\
                \alpha_1 &= \text{min} \; \left\{ 1, \frac{\mathcal{\hat{L}}(\bm{d}_{\text{obs}} | \bm{\hat{\theta}}')}{ \mathcal{\hat{L}}(\bm{d}_{\text{obs}} | \bm{\hat{\theta}}^{(j)})} \right\} & \text{(\textit{pCN})}
                \end{align*}
                \item If $u \sim U(0,1) > \alpha_1$ then set $\bm{\hat{\theta}}^{(j+1)} = \bm{\hat{\theta}}^{(j)}$ and return to (1); otherwise set $\bm{\hat{\theta}}^{(j+1)} = \bm{\hat{\theta}}'$ and continue to (4).
                \item If $t$ proposals have been accepted in the approximation subchain, continue to (5), otherwise return to  (1).
                \item  Given the latest realisation of the entire parameter set $\bm{\theta}^{(i)} = [\bm{\hat{\theta}}^{(i)}, \bm{\tilde{\theta}}^{(i)}]$ with fine parameters $\bm{\tilde{\theta}}^{(i)}$ and the transition kernel $q(\bm{\tilde{\theta}}' | \bm{\tilde{\theta}}^{(i)})$, generate a proposal for the fine parameters $\bm{\tilde{\theta}}'$ and set $\bm{\theta}' := [\bm{\hat{\theta}}', \bm{\tilde{\theta}}']$.
                \item Compute the likelihood ratio on the fine model between the proposal and the previous realisation: 
                \begin{equation*}
                \alpha_2 = \text{min} \left\{ 1, \frac{\pi_0(\bm{\theta}') \mathcal{L}(\bm{d}_{\text{obs}} | \bm{\theta}')}{\pi_0(\bm{\theta}^{(i)}) \mathcal{L}(\bm{d}_{\text{obs}} | \bm{\theta}^{(i)})} \frac{\pi_0(\bm{\hat{\theta}}^{(i)}) \mathcal{\hat{L}}(\bm{d}_{\text{obs}} | \bm{\hat{\theta}}^{(i)})}{\pi_0(\bm{\hat{\theta}}') \mathcal{\hat{L}}(\bm{d}_{\text{obs}} | \bm{\hat{\theta}}')} \right\}  \text{(\textit{AM})}
                \end{equation*}
                \begin{equation*}
                \alpha_2 = \text{min} \; \left\{ 1, \frac{\mathcal{L}(\bm{d}_{\text{obs}} | \bm{\theta}')}{\mathcal{L}(\bm{d}_{\text{obs}} | \bm{\theta}^{(i)})} 
                \frac{ \mathcal{\hat{L}}(\bm{d}_{\text{obs}} | \bm{\hat{\theta}}^{(i)})}{\mathcal{\hat{L}}(\bm{d}_{\text{obs}} | \bm{\hat{\theta}}')} \right\}  \text{(\textit{pCN})}
                \end{equation*} 
                \item If $u \sim U(0,1) > \alpha_2$ then set $\bm{\theta}^{(i+1)} = \bm{\theta}^{(i)}$, otherwise set $\bm{\theta}^{(i+1)} = \bm{\theta}'$.
            \end{enumerate}
    }}
\end{center}

\newpage
\subsection{Adaptive correction of the approximate posterior}
\begin{figure}[htbp]
\centering
\includegraphics[width=0.9\linewidth]{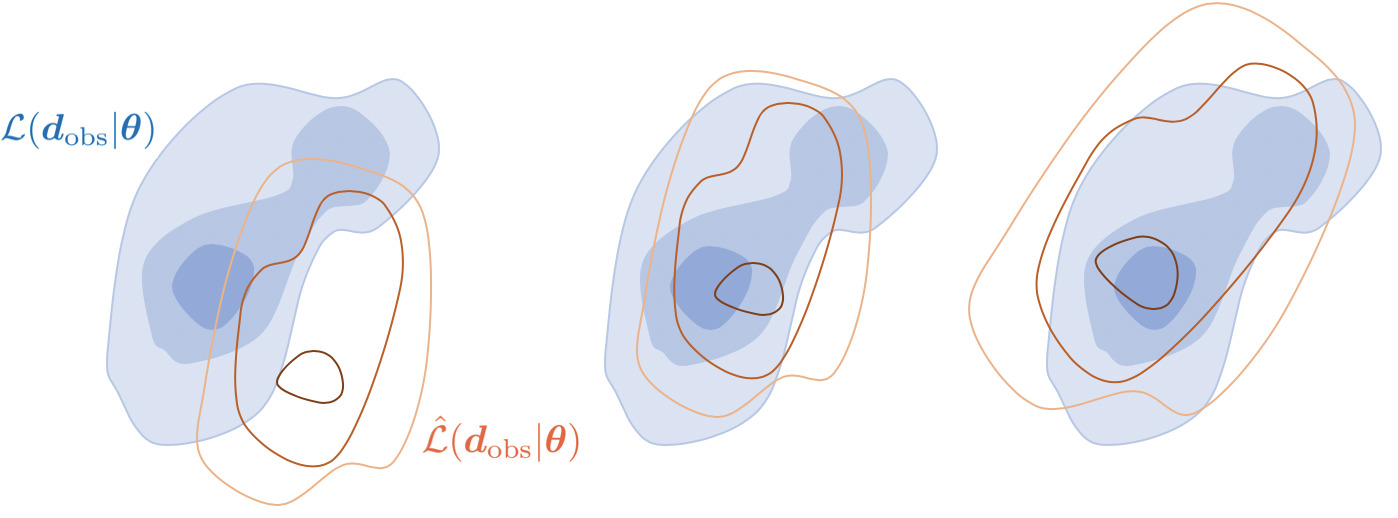}
\caption{Fine/target likelihood (blue) and approximate likelihood (red). (Left) Original likelihood before correction, (middle) corrected likelihood by a constant shift $\bm{\mu}_{\text{bias}}$ and (right) corrected approximate likelihood by multivariate Gaussian.}
\label{fig:inflation_adaption}
\end{figure}

Whilst in theory the modified delayed acceptance proposal described in Section~\ref{sec:DA}  will provide a convergent Metropolis-Hastings algorithm, there are cases in which the rate of convergence will be extremely slow. To demonstrate this, the left-hand contour plot in Fig.~\ref{fig:inflation_adaption} shows an artificially bad example. In this case the approximate model (red isolines) poorly captures the target likelihood distribution (blue density); there is a clear offset in the distributions, and the scale, shape and orientation of the approximate likelihood is incorrect. If using the modified delayed acceptance algorithm without alteration, it is easy to see that the proposal mechanism would struggle to traverse the whole of the target distribution, since much of it lies in the tails of the approximate likelihood distribution. As a result, in practice, we would observe extremely slow convergence to the true posterior; in practise -- at finite computational times -- results would contain a significant bias.

\smallskip
An ad hoc way to overcome this is to apply so-called {\em tempering} on the statistical model which drives the subchain. In this technique, the variance of the misfit ${\bm \Sigma}_\epsilon$ on the subchain is artificially inflated to capture the uncertainty in the approximate model. The issue in adopting this approach is the difficulty in selecting a robust inflation factor for tempering, particularly in higher dimensions. Furthermore, an isotropic inflation of the approximate posterior will in general be sub-optimal.

In this paper we instead implement an adaptive enhanced error model (EEM), which overcomes many of these challenges. Moreover, it is easy to implement and has negligible additional computational cost. Let $\mathcal{\hat{F}}$ denote the approximate forward map of the fine/target model $\mathcal{F}$. Then, following \cite{kaipio_statistical_2007,cui_posteriori_2018}, we apply a trick to the statistical model~\eqref{eq:inverse_problem} where we add and subtract the coarse map $\mathcal{\hat{F}}$. With some rearrangement we obtain the expression
\begin{equation} \label{eq:inverse_problem_bias}
\bm{d}_{\text{obs}} = \mathcal{F}(\bm{\theta}) + \bm{\epsilon} = \mathcal{F}(\bm{\theta}) + \mathcal{\hat{F}}(\bm{\theta}) - \mathcal{\hat{F}}(\bm{\theta}) + \bm{\epsilon} = \mathcal{\hat{F}}(\bm{\theta}) + \underbrace{\left(\mathcal{F}(\bm{\theta}) - \mathcal{\hat{F}}(\bm{\theta})\right)}_{:=\mathcal B({\bm \theta})} + \bm{\epsilon}.
\end{equation}
Here $\mathcal B(\bm{\theta}) = \mathcal{F}(\bm{\theta}) - \mathcal{\hat{F}}(\bm{\theta})$ is the bias associated with the approximation at given parameter values $\bm{\theta}$. We approximate this bias using a multivariate Gaussian distribution, i.e.  $\mathcal B \sim \mathcal N(\bm{\mu}_{\text{bias}}, \bm{\Sigma}_{\text{bias}})$, and therefore the likelihood function (\ref{eq:likelihood}) can be rewritten as
\begin{equation} \label{eq:likelihood_adaptive}
\mathcal{\hat{L}}(\bm{d}_{\text{obs}} | \bm{\theta}) = \exp \left (-\frac{1}{2} (\mathcal{\hat{F}}(\bm{\theta}) + \bm{\mu}_{\text{bias}} - \bm{d}_{\text{obs}})^\intercal (\bm{\Sigma}_{\text{bias}} + \bm{\Sigma}_e)^{-1} (\mathcal{\hat{F}}(\bm{\theta}) + \bm{\mu}_{\text{bias}} - \bm{d}_{\text{obs}}) \right).
\end{equation}
The influence of redefining the likelihood is best demonstrated geometrically, as shown in Fig. \ref{fig:inflation_adaption} (middle and right). Firstly, as shown in Fig. \ref{fig:inflation_adaption} (middle) we can make a better approximation by simply adding a shift of the mean bias $\bm{\mu}_{\text{bias}}$ to the original approximate model $\mathcal{\hat{F}}(\bm{\theta})$. This has the effect of aligning the `centre of mass' of each of the distributions. Secondly, we can learn the covariance structure of the bias. This has the effect of stretching and rotating the approximate distribution to give an even better overall approximation, as shown in Fig.~\ref{fig:inflation_adaption} (right). The final mismatch between the approximate and target distribution, will be driven by the assumption that bias can be represented by a multivariate Gaussian, although more complex distributions could be constructed using, for example, Gaussian process regression. Whilst this is an avenue to explore in the future, any such approach would surrender the simplicity of this approach, which from the results appears particularly effective.

\smallskip
The idea of using an EEM when dealing with model hierarchies originates from \cite{kaipio_statistical_2007}, who suggested to use samples from the prior distribution of parameters to construct the EEM prior to Bayesian inversion, so that
\begin{equation}
    \bm{\mu}_{\text{bias}}    = \frac{1}{N} \sum_{i=1}^{N} \mathcal{B}(\bm{\theta}^{(i)}) \quad \mbox{and} \quad    \bm{\Sigma}_{\text{bias}} = \frac{1}{N-1} \sum_{i=1}^{N} (\mathcal{B}(\bm{\theta}^{(i)}) - \bm{\mu}_{\text{bias}})(\mathcal{B}(\bm{\theta}^{(i)}) - \bm{\mu}_{\text{bias}})^{\intercal}
\end{equation}
The estimates for $\bm{\mu}_{\text{bias}}$ and $\bm{\Sigma}_{\text{bias}}$ could be obtained by sampling the prior distribution and comparing the approximate forward map against the target forward map. \ml{This approach has previously been successfully applied to a geophysical inverse problem by \cite{hansen_accounting_2014}, who compared the modelling error for a large number of crosshole tomography models}. \ml{However, since the model output generated by parameter sets drawn from the prior distribution may, on average, be biased significantly differently than samples drawn from a (relatively concentrated) posterior distribution, this approach may lead to an EEM that poorly represents the model bias associated with the posterior. If the approximate model is a good approximation \textit{on average}, constructing the EEM from the prior distribution would lead to an underestimation of the mean and an overestimation of the covariance of the bias, compared to an EEM constructed from the posterior.} Furthermore, in our example where the approximate model is built from samples from the prior, it is expected that such an approach would \ml{further} underestimate \ml{both the mean \textit{and} covariance of the bias}, since the neural network has been explicitly trained to minimise the error with respect to samples from the prior. 

\smallskip
Instead of estimating the bias using the prior, the posterior bias can be constructed on-line by iteratively updating its mean $\bm{\mu}_{\text{bias}}$ and covariance $\bm{\Sigma}_{\text{bias}}$ using coarse/fine solution pairs from the MCMC samples as suggested by \cite{cui_bayesian_2011}. \ml{Another similar approach was employed to a Bayesian geophysical problem by \cite{kopke_accounting_2018}, who collected model bias estimates while sampling, and used the bias estimates of the $k$-nearest-neighbours of each new coarse sample to construct a bias.} In this case we select
\begin{align}
    \bm{\mu}_{\text{bias},i+1}    &= \frac{1}{i+1} \big(i\bm{\mu}_{\text{bias},i} + \mathcal{B}(\bm{\theta}^{(i+1)}) \big) \quad \mbox{and} \\
    \bm{\Sigma}_{\text{bias},i+1} &= \frac{i-1}{i} \bm{\Sigma}_{\text{bias},i} + \frac{1}{i} (\mathcal{B}(\bm{\theta}^{(i+1)}) \: \mathcal{B}(\bm{\theta}^{(i+1)})^\intercal - \bm{\mu}_{\text{bias},i+1} \: \bm{\mu}_{\text{bias},i+1}^\intercal)
\end{align}
While this approach does not in theory guarantee ergodicity of the chain \ml{(as is also the case with the Adaptive Metropolis proposal)}, the bias distribution will converge as the chain progresses and adaptation diminishes, resulting in a \textit{de facto} ergodic process after an initial period of high adaptivity. This is a common feature of adaptive MCMC algorithms, as discussed in the classic paper on Adaptive Metropolis \cite{haario_adaptive_2001}. Our experiments showed that the bias distribution did indeed converge for every simulation, and that repeated experiments converged towards the same posterior bias distribution. Admitting a bias term in the inverse problem further introduces an issue of \textit{identifiability}, as highlighted in \cite{brynjarsdottir_learning_2014}. Since observations are now modelled as a sum of coarse model output and multiple stochastic terms, the stochastic terms $\mathcal{B} \sim N(\bm{\mu}_{\text{bias}}, \bm{\Sigma}_{\text{bias}})$ and $\bm{\epsilon} \sim N(0,\sigma^2 \mathbb{I}_n)$ are generally unidentifiable in the coarse model formulation\ml{, meaning that the bias $\mathcal{B}$ and the data modelling noise $\bm{\epsilon}$ are observationally equivalent, and not well-defined}.

\section{Results} \label{sec:results}
In this section, we examine the effectiveness of our proposed strategy on two synthetic groundwater flow problems: a two-dimensional problem with an isotropic covariance kernel and a three-dimensional problem with an anisotropic covariance kernel. For both examples, we begin by outlining the model setup, for which we select a `true' transmissivity field and a number of fixed observation points. For the first example, the influence of training size for the DNNs is examined, and the total cost of uncertainty quantification using a selection of DNNs is computed. For the second example we use a single DNN setup and analyse the resulting posterior marginal distributions and the quantity of interest. The first example was completed on commodity hardware -- an HP Elitebook 840 G5 with an Intel Xeon E3-1200 quad-core processor, while the second example was completed on a TYAN Thunder FT48T-B7105 GPU server with two Intel Xeon Gold 6252 processors and an NVIDIA RTX 2080Ti GPU.

\subsection{Example 1: 2D Unit Square}
\subsubsection{Model Setup}
This example was conducted on a unit square domain $\Omega=[0, 1]^2$, meshed using an unstructured triangular grid comprising \numprint{2601} degrees of freedom. Dirichlet boundary conditions were imposed on the left and right boundaries with hydraulic heads of 1 and 0, respectively. The top and bottom edges impose homogeneous no-flow Neumann boundary conditions. \ml{To avoid committing an inverse crime}, the covariance length scales of the ARD squared exponential kernel was set to \ml{$\bm{l} = (0.11, 0.11)^\intercal$ for data generation and $\bm{l} = (0.1, 0.1)^\intercal$ for the forward model used in sampling.} \ml{The chosen length scales} effectively result\ml{ed} in an isotropic covariance kernel, equal to the `classic' square exponential kernel with $l = 0.1$. This resulted in a KL decomposition with $>80\%$ of total signal energy in the 32 \ml{largest eigenvalues} and $>95\%$ of signal energy in the $64$ \ml{largest eigenvalues}. Hence, $32$ modes were included in the approximate model whilst $64$ modes where included in the fine model.

\begin{figure}[htbp]
    \centering
    \begin{subfigure}{0.4\textwidth}
        \centering
        \includegraphics[width=1\linewidth]{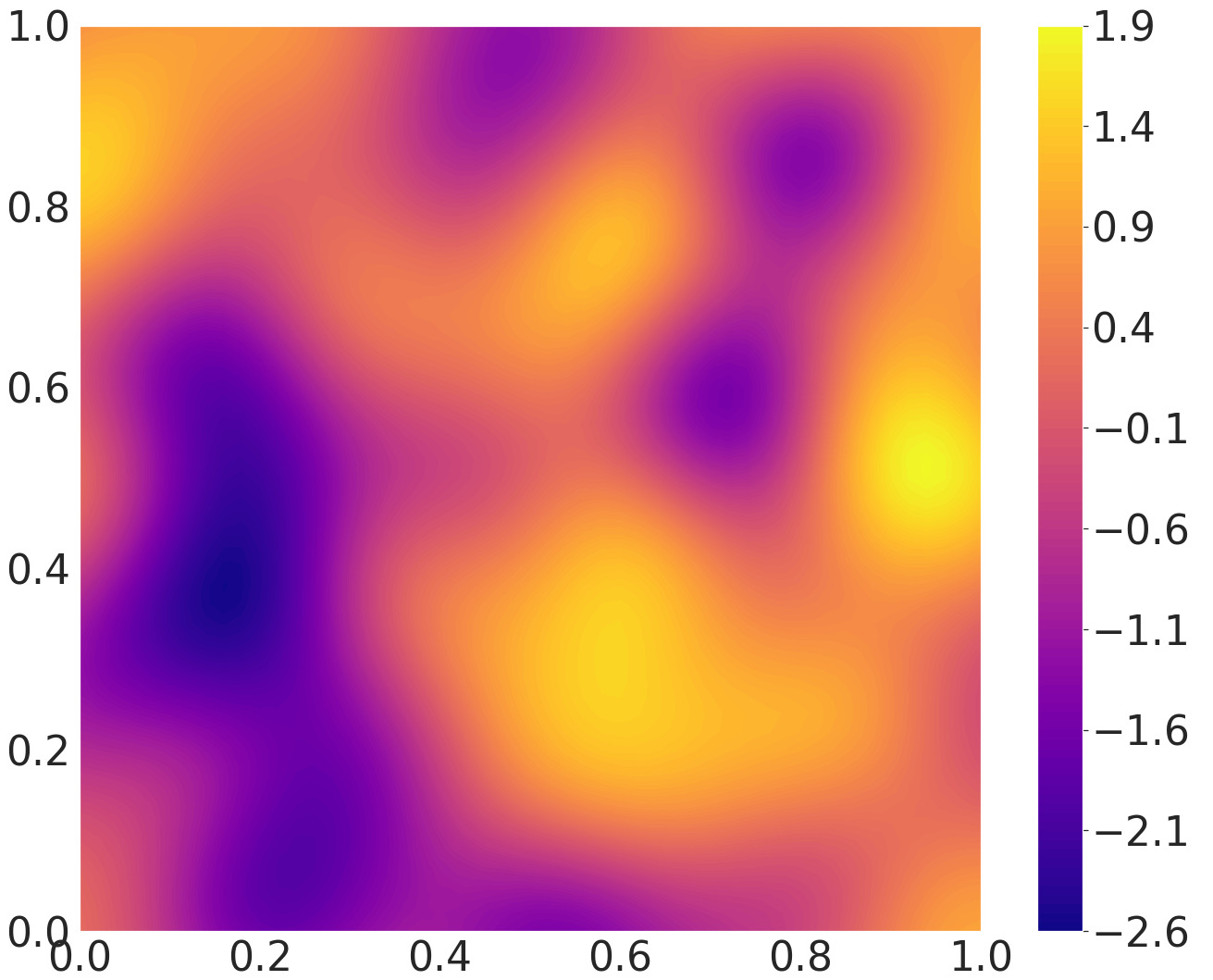}
        \caption{Log-Transmissivity.}
        \label{fig:true_field}
    \end{subfigure}
    \hspace{0.5cm}
    \begin{subfigure}{0.4\textwidth}
        \centering
        \includegraphics[width=1\linewidth]{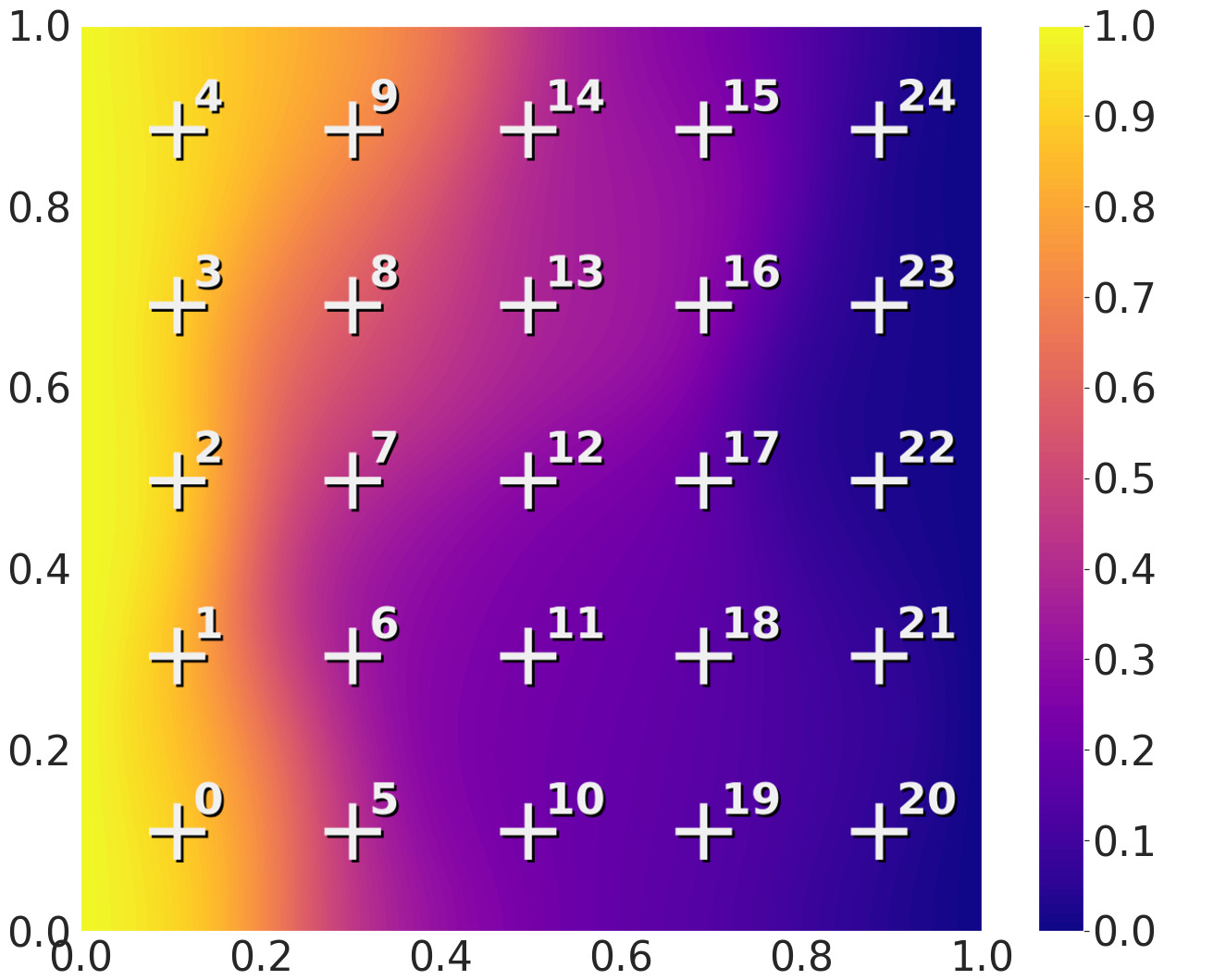}
        \caption{Hydraulic head.}
        \label{fig:true_solution}
    \end{subfigure}
    \caption{``True'' transmissivity field, its corresponding solution and sampling points.}
    \label{fig:ground_truth}
\end{figure}

Figure~\ref{fig:true_field} shows the `true' transmissivity field that we attempt to recover through our MCMC methodology and the modelled, corresponding hydraulic head. Synthetic samples for the likelihood function were extracted at 25 points on a regular grid with a horizontal and vertical spacing of 0.2m (Figure \ref{fig:true_solution}), \ml{and these data were pertubated with white noise with covariance $\bm{\Sigma}_e = 0.001 \: \mathbb{I}_m$}.

\subsubsection{Deep Neural Network Design, Training and Evaluation}
We evaluated a selection of different DNNs to investigate the impact of various network depths and activation functions on the DNN performance. Table \ref{tab:ann_design} shows the layers of the employed DNNs, the number of nodes in each layer and their corresponding activation functions. DNN1 and DNN2 had three hidden layers, while DNN3 and DNN4 had only two, as the ReLU layer with $8k$ nodes was not included in these networks. The output layer of DNN1 and DNN3 consisted of nodes with an exponential activation function $E(x) = e^x$, resulting in a strictly positive output. The DNNs with an exponential activation function in the final layer tended overall to lead to the best performance.

\begin{table}[htbp]
    \centering
    \footnotesize
    \caption{Neural network layers and activation functions in the model approximation DNNs.} 
    \begin{tabular}{llllll}
        \toprule
        Layer & \# Nodes                        & \multicolumn{4}{c}{Activation Functions} \\ \cmidrule{3-6}
        &                                          & DNN1         & DNN2         & DNN3         & DNN4    \\
        \midrule
        Input & $k$ KL coefficients                & ---          & ---          & ---          & ---        \\
        1     & $4k$                               & Sigmoid      & Sigmoid      & Sigmoid      & Sigmoid \\
        2     & $8k$                               & ReLU         & ReLU         & ---          & ---     \\
        3     & $4k$                               & ReLU         & ReLU         & ReLU         & ReLU    \\
        Output & $m$ datapoints                    & Exponential  & Linear       & Exponential  & Linear  \\
        \bottomrule
    \end{tabular}
    \label{tab:ann_design}
\end{table}

Each DNN was trained on a set of samples from the prior distribution of parameters $\pi_0(\bm{\theta}) = \mathcal{N}(0,\mathbb{I}_k)$, in advance of running the MCMC. Hence, the DNN samples were drawn from a Latin Hypercube \cite{mckay_sampling_1979} in the interval $[0,1]$ and transformed to the standard normal distribution using the \textit{probit}-function, such that $\bm{\theta}_{train} \sim \mathcal{N}(0,\mathbb{I}_k)$. The \ml{coarse, 32-mode} FEM model was then run for every parameter sample, obtaining for each a vector of model outputs at sampling points given parameters. We trained and tested each DNN on a range of different sample sizes, namely $N_{\text{DNN}} = \{2000, 4000, 8000, 16000, 32000, 64000\}$, where $N_{\text{DNN}} = N_{train}+N_{test}$, with a 9:1 training/test splitting ratio.  Each DNN was then trained for 200 epochs with a batch size of 50 using the \texttt{rmsprop} optimiser  \cite{rmsprop}. 

\begin{figure}[htbp]
    \centering
    \includegraphics[width=0.6\textwidth]{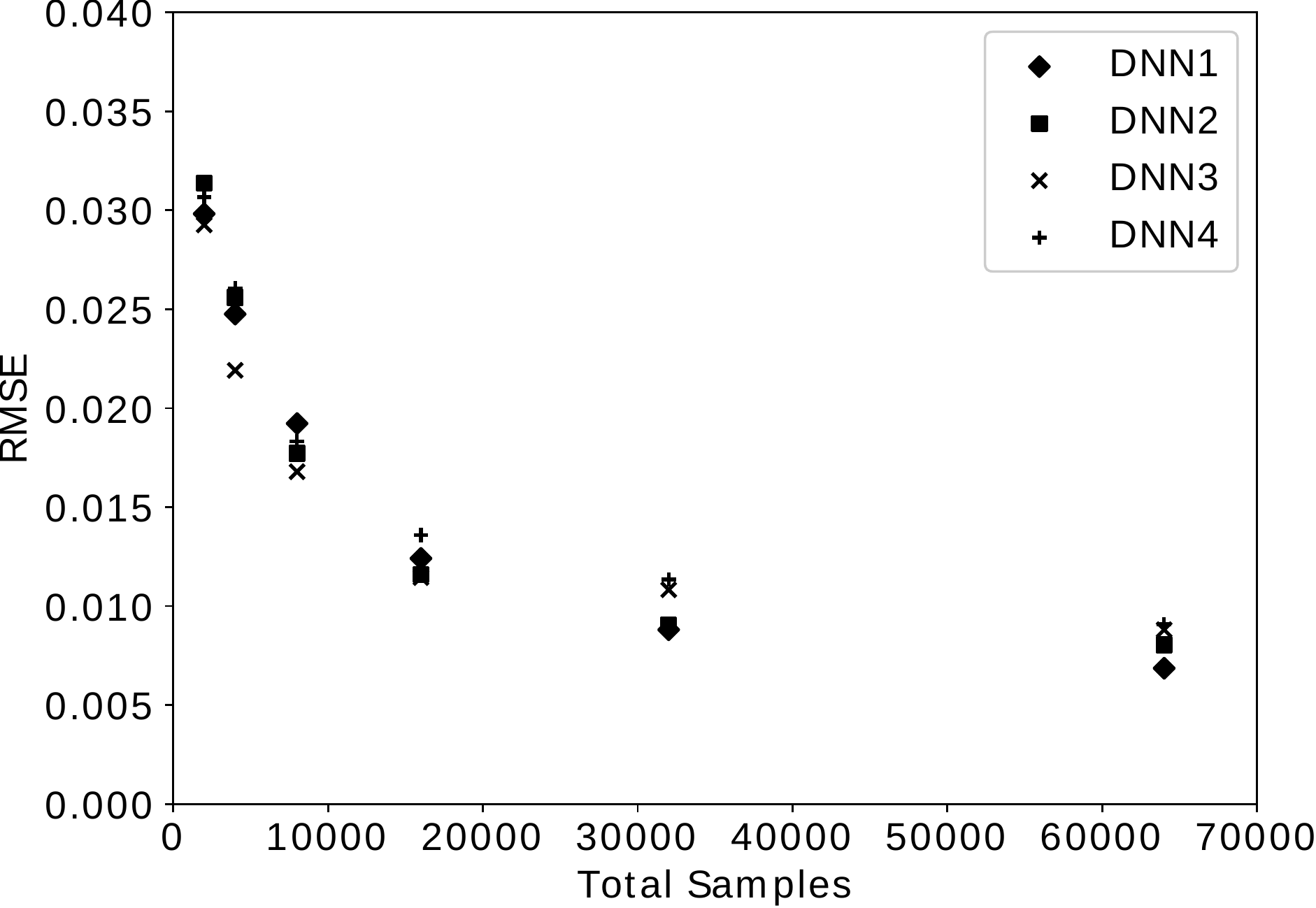}
    \caption{Testing performance (RMSE) of each DNN against the total sample size ($N_{\text{DNN}} = N_{train}+N_{test}$). Please refer to table \ref{tab:ann_design} for details in the structure of each DNN.}
    \label{fig:multi_mlp_performance}
\end{figure}

\smallskip
Deep Neural Networks performance was compared using the RMSE of their respective testing dataset
\begin{equation}\label{eqn:RMSE}
\text{RMSE} = \sqrt{\frac{1}{n} \sum_{i = 1}^{n} (h_i - \hat{h}_i)^2}
\end{equation}
The residual RMSE \eqref{eqn:RMSE} of each DNN was computed to compare the network designs described in Table \ref{tab:ann_design} and to investigate the influence of training dataset size on the DNN performance (Figure~\ref{fig:multi_mlp_performance}). As expected, each DNN performed better as the number of samples in the training dataset were increased. In comparison, the DNN design had much less influence on the testing performance, suggesting that the main driver for constructing an accurate surrogate model, within the bounds of the examined DNN designs, was the number of training samples. For the remaining analysis, we chose the  \ml{network design resulting in the overall lowest RMSE at $N_{\text{DNN}} = 64000$}, namely DNN1\ml{, and the sample sizes} $N_{\text{DNN}} = \{4000, 16000, 64000\}$. 

\begin{figure}[h!]
    \centering
    \includegraphics[width=0.9\textwidth]{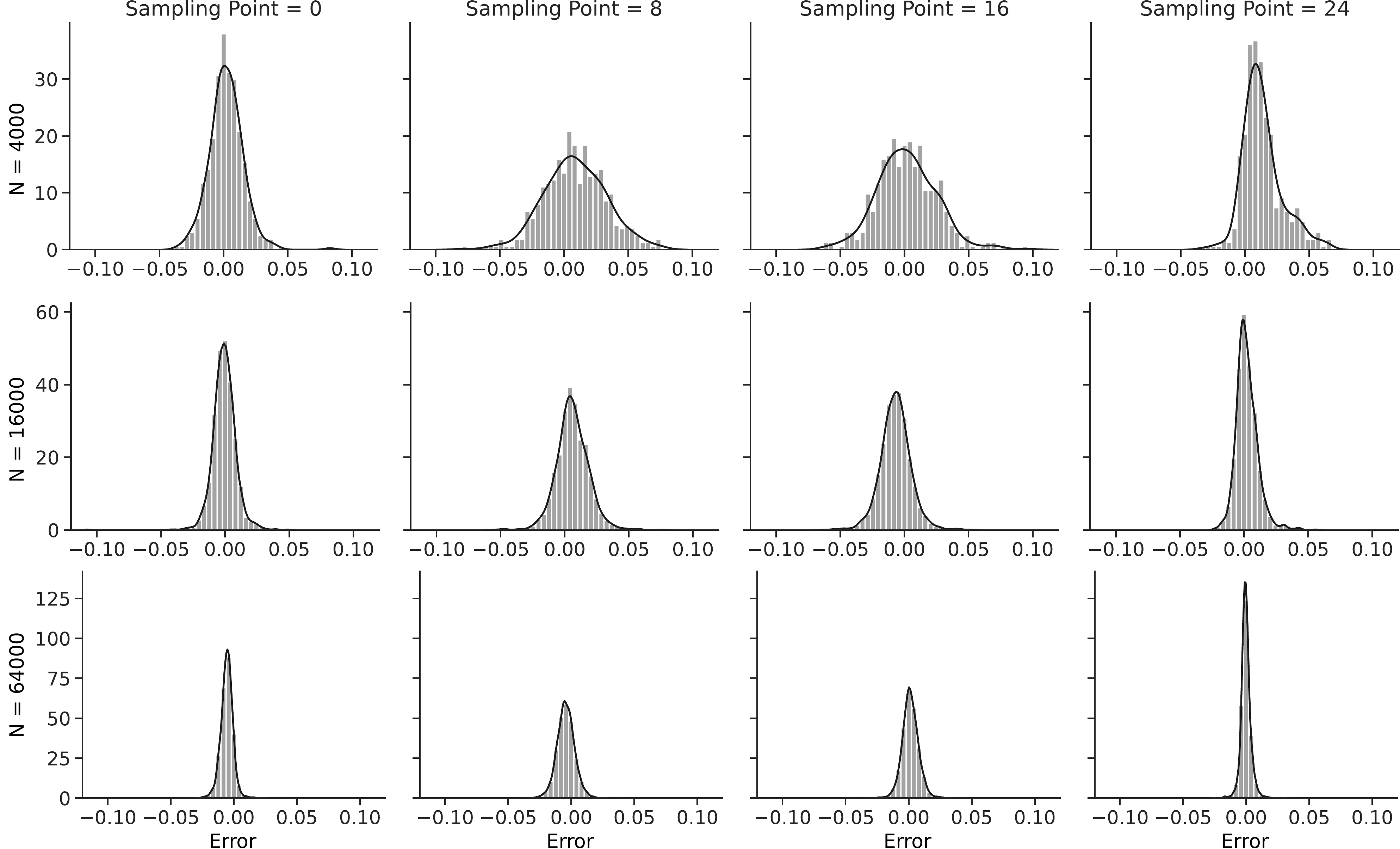}
    \caption{Density plot of the error ($e=h_{\text{true}} - h_{\text{pred}}$) of the testing dataset for DNN1 trained and tested on $N_{\text{DNN}} = \{4000, 16000, 64000\}$ samples, for sampling points 0, 8, 16 and 24. Bars show density of each bin, while the curve shows Gaussian kernel density estimate.}
    \label{fig:error_test}
\end{figure}

Further performance analysis consisted of analysing the DNN error $e=h_{\text{true}} - h_{\text{pred}}$ for true and predicted heads ($h_{\text{true}}$ and $h_{\text{pred}}$, respectively) for datapoints 0, 8, 16 and 24.  (Figure \ref{fig:error_test}). All error distributions were approximately Gaussian, with the errors for the DNN with $N_{\text{DNN}} = 4000$ exhibiting some right skew at sampling point 24. For all DNNs, the sampling points closer to the boundaries (at sampling points 0 and 24) had lower errors than those further away, since the heads close to the boundaries were more constrained by the model.

\subsubsection{Uncertainty Quantification}
For inversion and uncertainty quantification, we chose a multivariate standard normal distribution as the prior parameter distribution, $\pi_0(\bm{\theta}) = \mathcal{N}(0,\mathbb{I}_k)$ and set the error covariance to $\bm{\Sigma}_e = 0.001 \: \mathbb{I}_m$. While computationally convenient, the zero-centred prior in practice favours transmissivity field realisations capable of reproducing the observed heads with as little variation as possible. In total, \ml{eight} different sampling strategies were investigated, namely single level `Vanilla' MCMC, \ml{with no delayed acceptance, no adaptivity, and using only the 64-mode fine model}; DA using three different DNNs trained and tested on $N_{\text{DNN}} = \{4000, 16000, 64000\}$ samples \ml{as the coarse model and the 64-mode model as the fine}; and DA with an enhanced error model (DA/EEM) using the same three DNNs. \ml{The offset length $t$ for the DA strategies was manually tuned to achieve an acceptance rate of $a \in [0.2, 0.4]$. To investigate the effect of the offset length $t$ independently of other factors, an additional simulation with $N_{\text{DNN}} = 64000$ and $t=1$ was also completed.} \ml{In this first example,} every simulation was completed using the pCN transition kernel, \ml{with $\beta = 0.15$}. Each MCMC sampling strategy was repeated ($n=32$) using randomly generated random seeds, to ensure that every starting point would converge towards the same stationary distribution and to allow for cross-chain statistics to be computed. Results given in this section pertain to these multi-chain samples rather than individual MCMC realisations, unless otherwise stated.

\begin{figure}[htbp]
    \centering
    \begin{subfigure}{0.4\textwidth}
        \centering
        \includegraphics[width=1\linewidth]{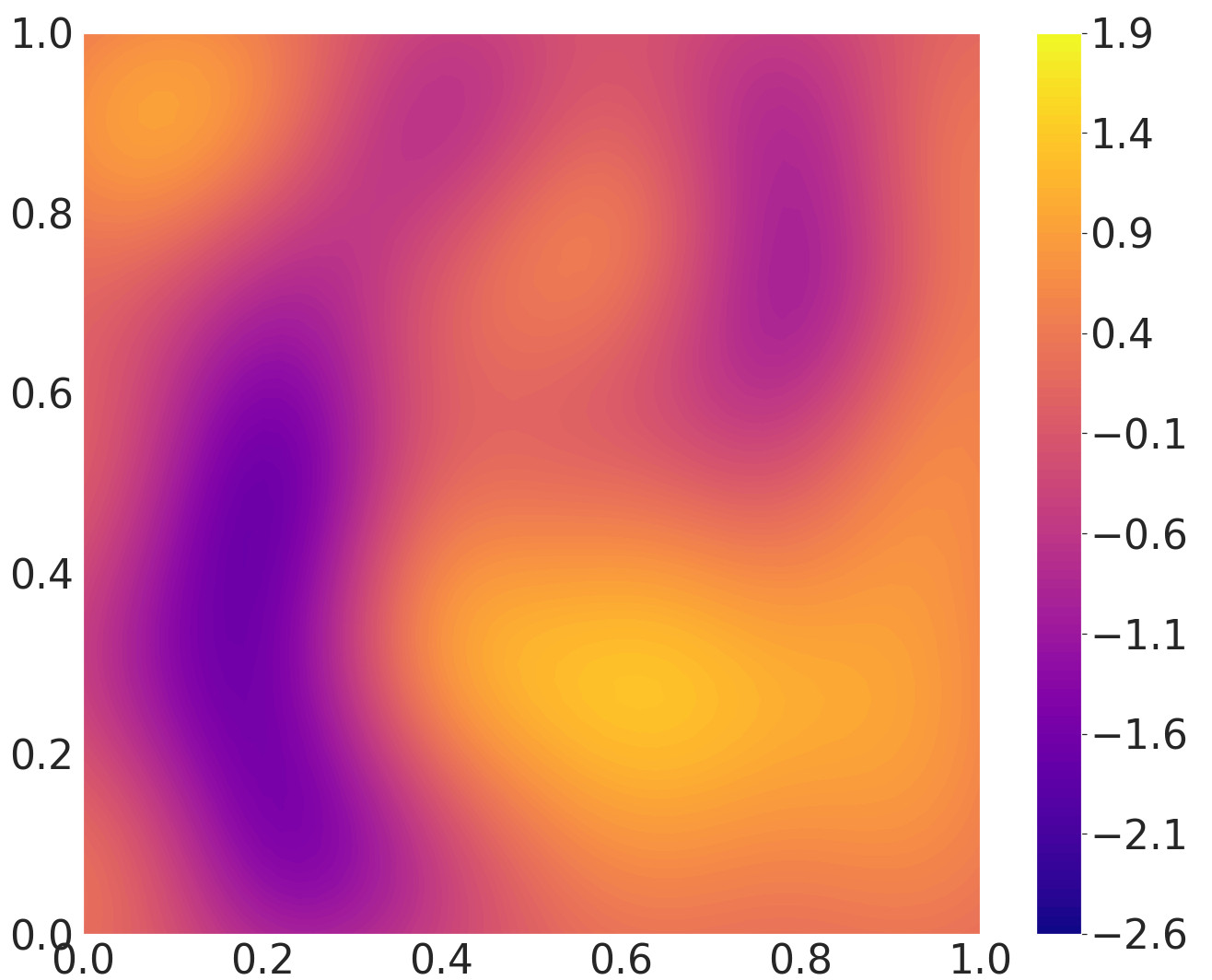}
        \caption{Mean of recovered log-transmissivity for Vanilla pCN sampler.}
        \label{fig:recovered_mean_reference}
    \end{subfigure}
    \hspace{0.5cm}
    \begin{subfigure}{0.4\textwidth}
        \centering
        \includegraphics[width=1\linewidth]{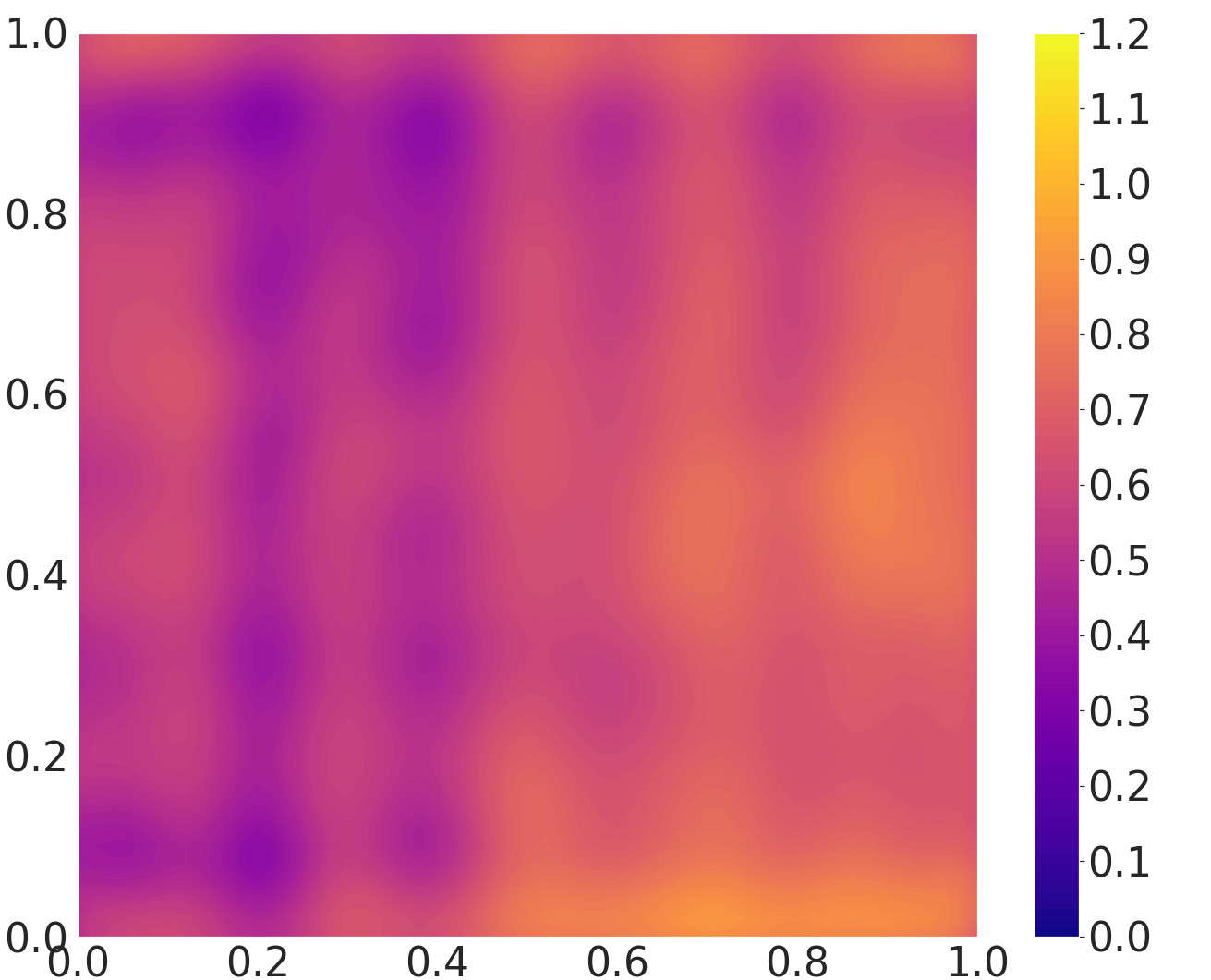}
        \caption{Variance of recovered log-transmissivity for Vanilla pCN sampler.}
        \label{fig:recovered_variance_reference}
    \end{subfigure}
    ~
    \begin{subfigure}{0.4\textwidth}
        \centering
        \includegraphics[width=1\linewidth]{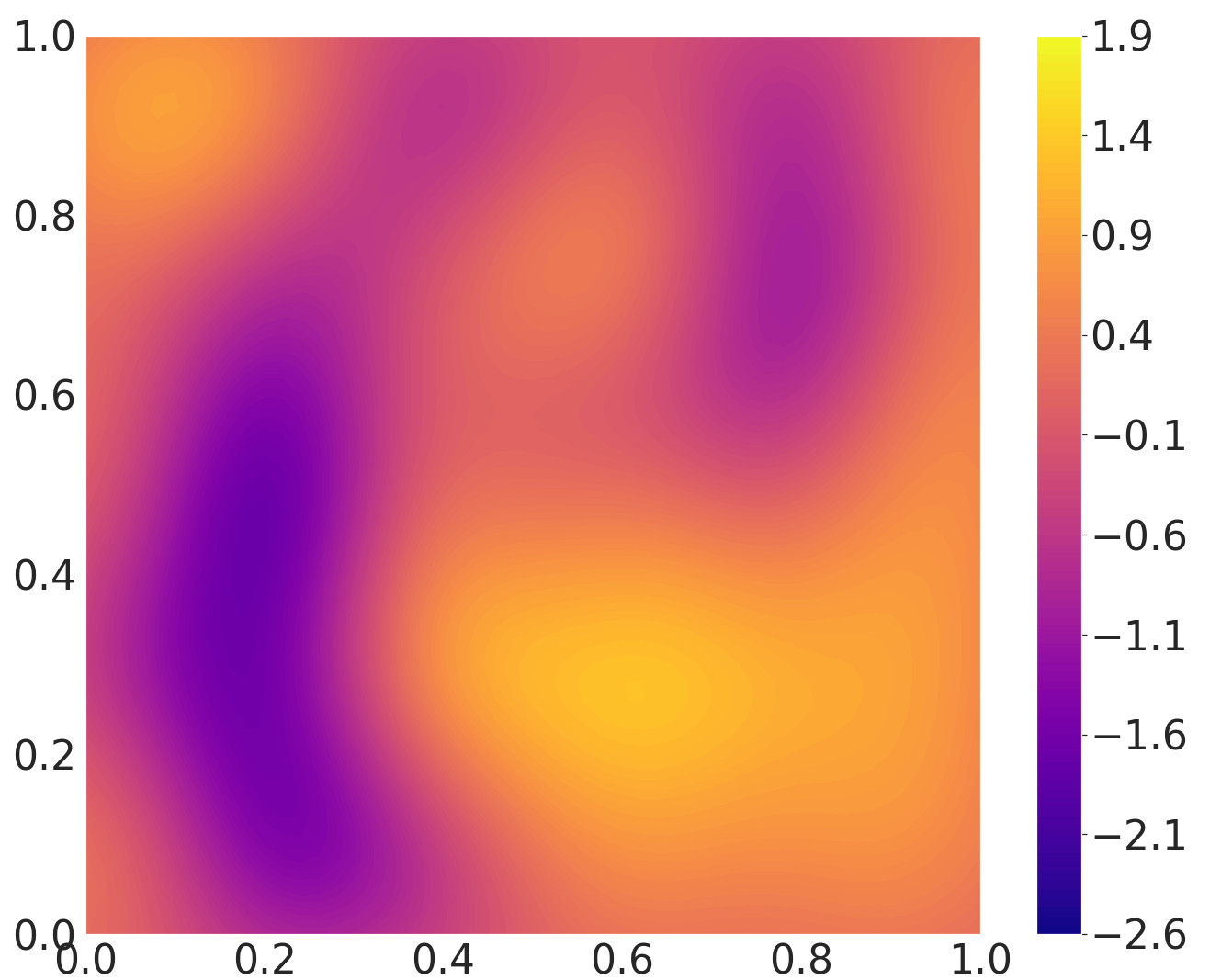}
        \caption{Mean of recovered log-transmissivity for DA/EEM, $N_{DNN} = 16000$.}
        \label{fig:recovered_mean_dnn_16000_eem}
    \end{subfigure}
    \hspace{0.5cm}
    \begin{subfigure}{0.4\textwidth}
        \centering
        \includegraphics[width=1\linewidth]{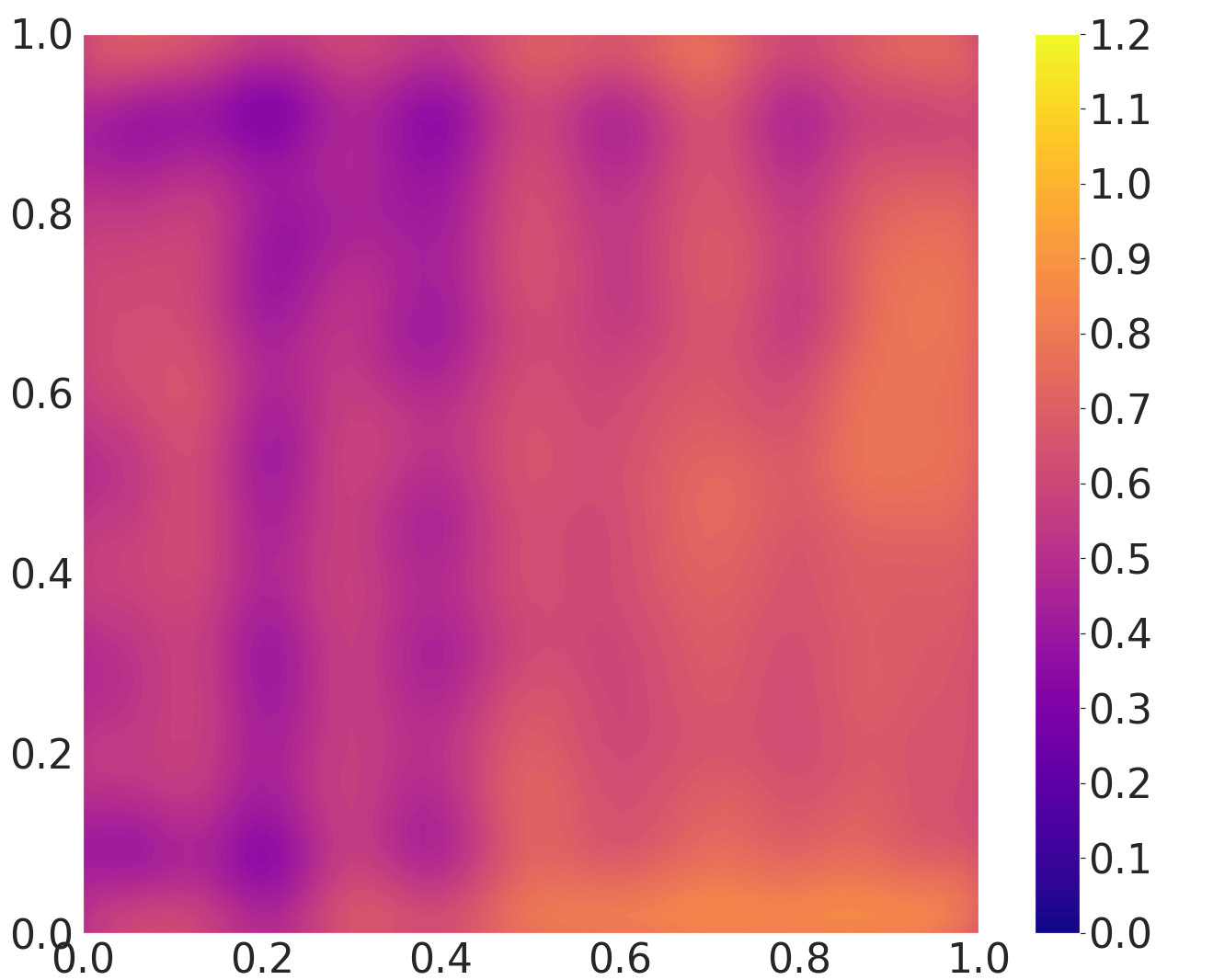}
        \caption{Variance of recovered log-transmissivity for DA/EEM, $N_{DNN} = 16000$.}
        \label{fig:recovered_variance_dnn_16000_eem}
    \end{subfigure}
    \caption{\ml{Mean and variance ($n=32$) of recovered log-transmissivity fields using Vanilla pCN sampler (top) and DA/EEM MCMC with $N_{\text{DNN}} = 16000$ (bottom). Please find corresponding plots of every sampling strategy in \ref{app:recovered}}}
    \label{fig:recovered}
\end{figure}

Our sampling strategies recovered the ground truth with good accuracy. Figure \ref{fig:recovered} shows the mean and variance of the recovered field from the DA/EEM MCMC using the DNN with $N_{\text{DNN}} = 64000$. All recovered fields exhibit higher smoothness than the ground truth, which \ml{can} be attributed to the relatively low number of sampling points and their regular distribution on the domain\ml{, in combination with the regularisation introduced by the prior. Since the KL decomposition incorporated $>95\%$ of the signal energy, the truncation would have contributed only marginally to the smoothing}. None of the chains recovered the local peak in transmissivity on the right side of the domain, since there was too little data to discover this particular feature. \ml{However, this peak is clearly encapsulated by the posterior variance, as shown in Figure \ref{fig:recovered_variance_reference} and \ref{fig:recovered_variance_dnn_16000_eem}.}

While the recovered fields indicate that every MCMC sampling strategy converged towards the desired stationary distribution, they do not reveal the relative efficiency of each strategy. Hence, the Effective Sample Size ($N_{eff}$) was computed for each MCMC realisation. Every DA sampling strategy produced higher $N_{eff}$ than the Vanilla pCN sampler, relative to the simulation time, with a clear correlation between DNN testing performance and $N_{eff}$. This was mainly because the better performing DNNs allowed for a longer coarse chain offset without diverging. Moreover, utilising the EEM produced even higher $N_{eff}$ for every DA chain (Table \ref{tab:results}).

\begin{table}[htbp]
    \centering
    \footnotesize
    \caption{Results for various MCMC sampling strategies, means of multiple chains with $n=32$. $N_{\text{DNN}}$ is the number of total samples used to construct the DNN. $t$ is the improved DA offset length. $N_C$ / $N_F$ is the final length of the coarse and fine chain, respectively, after subtracting burnin. \textit{Acc. rate} is the fine chain acceptance rate. \textit{Time (min)} is the total running time of the simulation in minutes. $N_{eff}$ is the Effective Sample Size.}
    \begin{tabular}{lllrrrrrr}
        \toprule
        Strategy     & N\textsubscript{DNN}   & $t$ & $N_C$ / $N_F$    & Acc. Rate 	& Time (min)     & $N_{eff}$      \\
        \midrule
        \ml{Vanilla} & ---         & ---      & --- / \ml{40000}       & \ml{0.33}  & \ml{32.1}     & \ml{85.6}   \\
        DA           & 4000        & 2        & 85461.9 / 20000        & 0.27       & 16.2          & 64.5  \\
        DA/EEM       & 4000        & 2        & 78853.4 / 20000        & 0.31       & 15.2          & 79.0  \\
        DA           & 16000       & 4        & 172383.1 / 20000       & 0.27       & 18.2          & 116.3  \\
        DA/EEM       & 16000       & 4        & 178978.4 / 20000       & 0.30       & 18.4          & 143.6  \\
        DA           & 64000       & 8        & 336447.5 / 20000       & 0.24       & 30.1          & 196.5 \\
        DA/EEM       & 64000       & 8        & 377524.4 / 20000       & 0.30       & 29.9          & 235.7 \\
        \ml{DA/EEM}  & \ml{64000}  & \ml{1}   & \ml{56824.3 / 20000}   & \ml{0.57}  & \ml{15.3}    & \ml{68.6} \\
        \bottomrule
    \end{tabular} 
    \label{tab:results}
\end{table}

\subsubsection{Total cost}
\begin{figure}[htbp]
        \centering
        \begin{subfigure}{0.4\textwidth}
            \centering
            \includegraphics[width=1\linewidth]{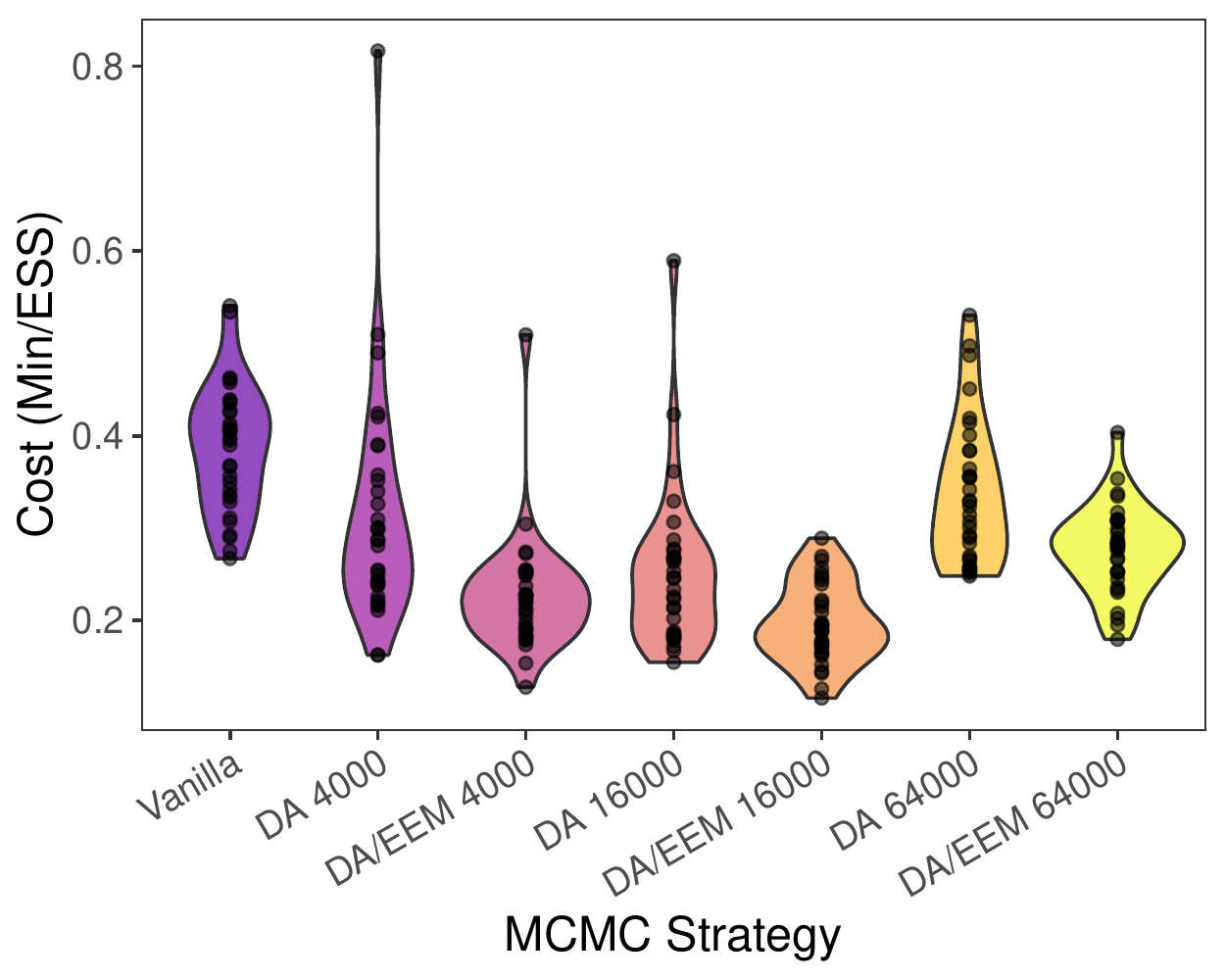}
            \caption{\ml{Total cost (conservative) with the full cost of constructing the DNN factored into all independent DA chains.}}
            \label{fig:cost_conservative}
        \end{subfigure}
        \hspace{0.5cm}
        \begin{subfigure}{0.4\textwidth}
            \centering
            \includegraphics[width=1\linewidth]{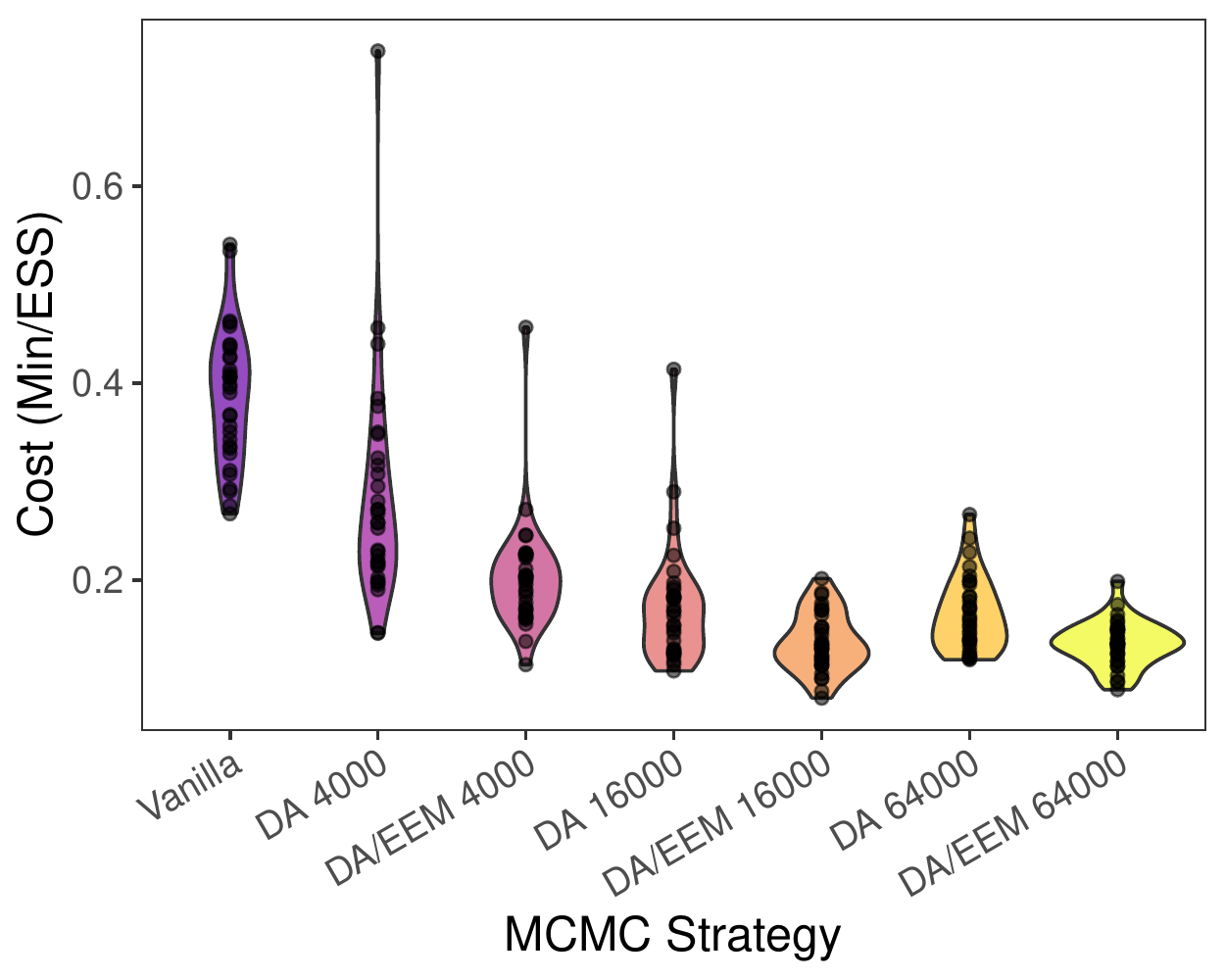}
            \caption{\ml{Total cost (normalised) with the cost of constructing the DNN distributed between independent DA chains.}}
            \label{fig:cost_normalised}
        \end{subfigure}
        \caption{Violinplots showing the total cost $\text{C}_{\text{total}}$ of each MCMC strategy with $n=32$. Points show independent Markov Chains.}
    \label{fig:cost}
\end{figure}

Since the DA chains required computation of a significant number of fine model solutions and training of a DNN in advance of running the chain, the total cost $\text{C}_{\text{total}}$ of each strategy was computed as 
\begin{equation} \label{eq:cost}
\text{C}_{\text{total}} = \frac{t_{\text{fine}} + t_{\text{train}} + t_{\text{run}}}{N_{eff}}
\end{equation}
where $t_{\text{fine}}$ was the time spent on precomputing fine model solution, $t_{\text{train}}$ was the time spent on training the respective DNN, $t_{\text{run}}$ was the time taken to run the chain and $N_{eff}$ was the resulting effective sample size (Fig. \ref{fig:cost}).

The mean cost of every DA chain was lower than that of the Vanilla pCN chain, with the chains using the EEM consistently cheaper than their non-EEM counterparts. Moreover, using the EEM reduced the variance of the cost in repeated experiments, allowing each repetition to produce a consistently high $N_{eff}$. The overall cheapest inversion was completed using the DNN trained on \numprint{16000} samples using the EEM, reducing the total cost, relative to the Vanilla pCN MCMC, with \ml{50}\%. Notice that these results are extremely conservative in the sense that the entire cost of evaluating every DNN training sample and training the DNN in serial on a CPU was factored into the cost of every repetition, even though the same DNN was used for all the repetitions within each sampling strategy. The precomputation cost can be dramatically reduced by evaluating the DNN samples in parallel and utilising high-performance hardware, such as GPUs, for training the DNN.

\subsection{Example 2: 3D Rectangular Cuboid}
\subsubsection{Model Setup}
\begin{figure}[h]
    \centering
    \begin{subfigure}{0.85\textwidth}
        \centering
        \includegraphics[width=0.9\linewidth]{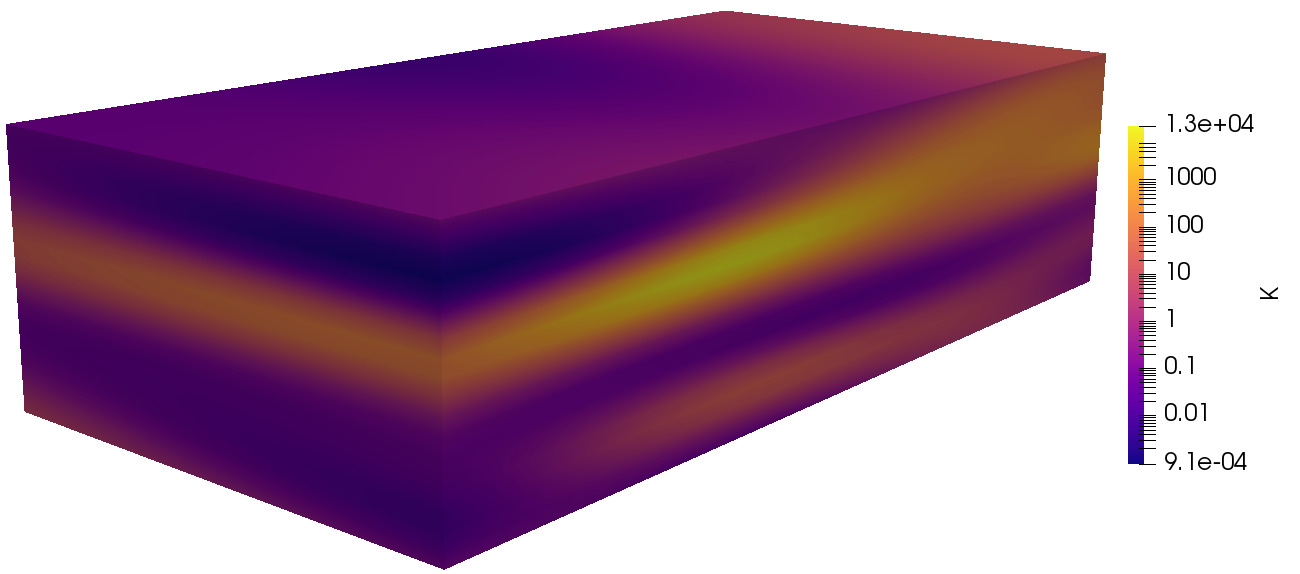}
        \caption{Log-Conductivity of ground truth.}
        \label{fig:true_field_3d}
    \end{subfigure}
    ~
    \begin{subfigure}{0.85\textwidth}
        \centering
        \includegraphics[width=0.9\linewidth]{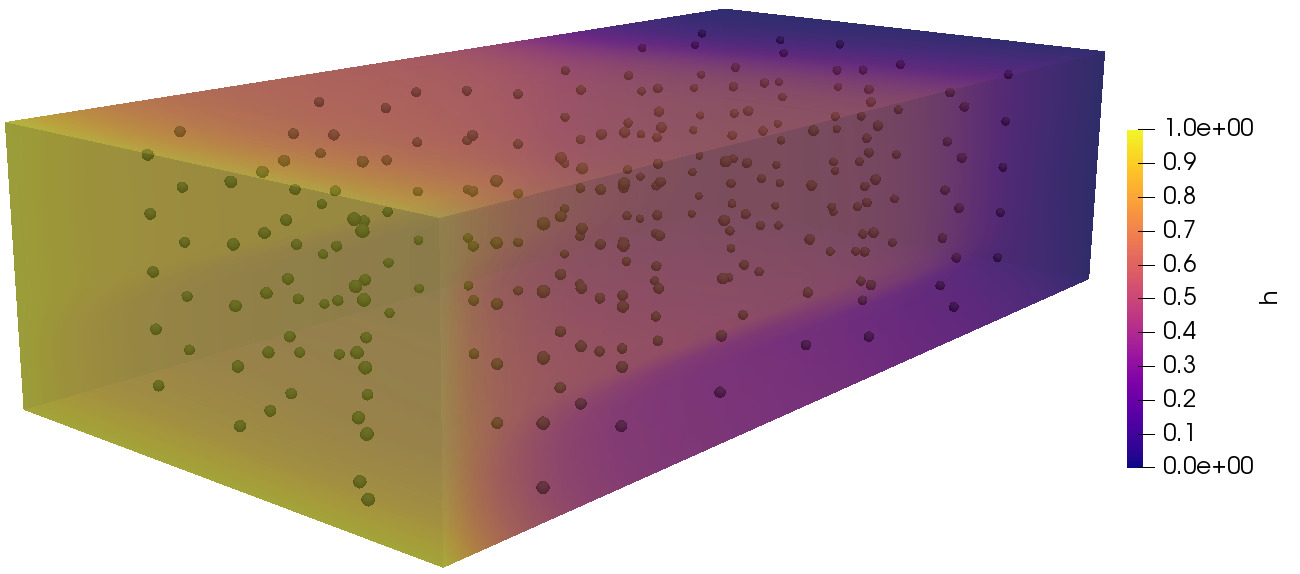}
        \caption{Hydraulic head of ground truth and location of sampling points.}
        \label{fig:true_head_3d}
    \end{subfigure}
    \caption{``True'' conductivity field, its corresponding solution and sampling points.}
    \label{fig:ground_truth_3d}
\end{figure}

This example was conducted on a rectangular cuboid domain $\Omega=[0,2] \times [0,1] \times [0,0.5]$ meshed using an unstructured tetrahedral grid with 10,416 degrees of freedom (Figure \ref{fig:ground_truth_3d}). Dirichlet boundary conditions of $h = 1$ and $h = 0$ were imposed at $x_1 = 0$ and $x_1 = 2$, respectively. No-flow Neumann conditions were imposed on all remaining boundaries.

The covariance lengths scales for ARD squared exponential covariance kernel were set to $\bm{l} = (0.55, 0.95, 0.06)^\intercal$ \ml{for data generation and $\bm{l} = (0.5, 1.0, 0.05)^\intercal$ for the forward model used in sampling}, resulting in a highly anisotropic random process with high variation in the $x_3$ direction to simulate geological stratification, some variation in the $x_1$ direction and little variation in the $x_2$ direction (Figure \ref{fig:true_field_3d}). Like in the first model, the random process was truncated at 64 KL eigenmodes for the fine model and 32 KL eigenmodes for the coarse model, embodying $>97\%$ and $>90\%$ of the total signal energy, respectively.

We drew $w = 50$ sampling well locations randomly using the Maximin Latin Hypercube Design \cite{morris_exploratory_1995}, and samples of hydraulic head were extracted at each well at datums $x_3 = \{0.05, 0.15, 0.25, 0.35, 0.45\}$, measured from the bottom of the domain, resulting in $m = 250$ datapoints in total (Figure \ref{fig:true_head_3d}). \ml{These data were pertubated with white noise with covariance $\bm{\Sigma}_e = 0.001 \: \mathbb{I}_m$}.

For this example, we first converged the conductivity parameters to the Maximum a posteriori (MAP) estimate $\bm{\theta}_{MAP} =  \underset{\theta}{\arg\max} \: \pi_0(\bm{\theta}) \mathcal{L}(\bm{d}_{\text{obs}} | \bm{\theta})$ using gradient descent, since initial MCMC experiments struggled to converge to the posterior distribution for random initial parameter sets.

\subsubsection{Deep Neural Network Design, Training and Evaluation}
Training a DNN to accurately emulate the model response for this setup was challenging, and we found no single combination of neural network layers and activation functions that would predict the head at every datapoint with sufficient accuracy. We hypothesise that this limitation could be caused by a strong ill-posedness of the DNN -- for a single neural network, the output dimension greatly exceeded the input dimension, \ml{i.e.} $m >> k$ \ml{where $m = 250$ is the number of datapoints, and $k = 32$ is the coarse model KL modes}.
\begin{figure}[htbp]
    \centering
    \includegraphics[width=0.4\textwidth]{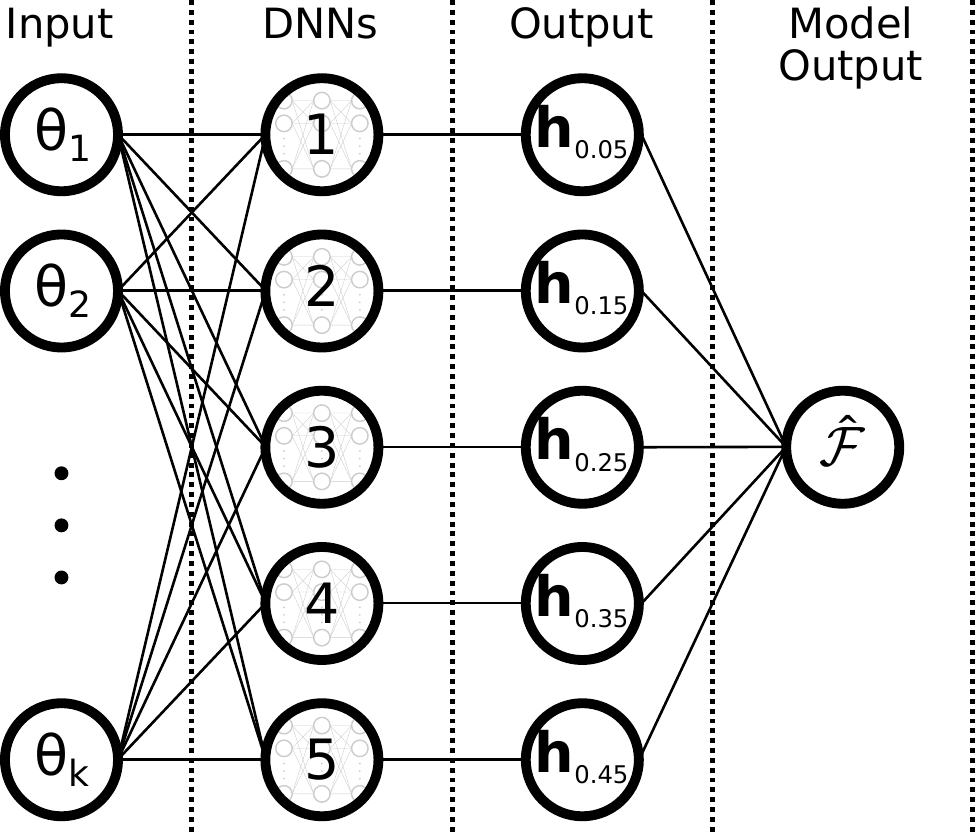}
    \caption{Layout of the multi-DNN design. Each DNN outputs a vector $\bm{h}_{x_3}$ vector of $w$ head predictions at datum $x_3$.}
    \label{fig:multi_mlp_3d}
\end{figure}
When we instead predicted the heads at each datapoint datum using a separate DNN, we found that we could utilise largely the same DNN design as had been employed in the first example. Hence, to predict the head at all datapoints, we utilised five identically designed but independent DNNs (Figure \ref{fig:multi_mlp_3d}), each with four hidden layers and activation functions as indicated in Table \ref{tab:ann_design_3d}.
\begin{table}[htbp]
    \centering
    \footnotesize
    \caption{Layers and activation functions in the four DNNs. Each DNN takes all $k$ KL coefficients as input and predicts the head $\bm{h}_{x_3}$ at $w$ wells for a given datum.}
    \begin{tabular}{llllll}
        \toprule
        Layer & \# Nodes                        & Activation Functions \\
        \midrule
        Input & $k$ KL coefficients             & ---           \\
        1     & $4k$                            & Sigmoid       \\
        2     & $8k$                            & ReLU          \\
        3     & $8k$                            & ReLU          \\
        3     & $4k$                            & ReLU          \\
        Output & $w$ wells                      & Exponential        \\
        \bottomrule
    \end{tabular}
    \label{tab:ann_design_3d}
\end{table}
Each DNN was trained and tested on a dataset of $N_{DNN} = 16000$ samples with KL coefficients drawn from a Latin Hypercube \cite{mckay_sampling_1979} in the interval $[0,1]$ and transformed to a normal distribution centered on the MAP estimate of the parameters $\bm{\theta}_{MAP}$, i.e. $\bm{\theta}_{train} \sim \mathcal{N}(\bm{\theta}_{MAP},\mathbb{I}_k)$. This was done to increase the density of samples and thus improve the DNN prediction at and around the MAP point, which ideally equals the mode of the posterior distribution. The DNNs were then trained for 200 epochs using a batch size of 50, the MSE loss function and the \texttt{rmsprop} optimiser \cite{rmsprop}. Figure \ref{fig:multi_dnn_performance} shows performance plots of each DNN for both the training (top\ml{)} and the testing (bottom) datasets. While every DNN is clearly moderately biased by the training data, they all performed adequately with respect to the testing data.

\begin{figure}[htbp]
    \centering
    \vspace{0.2cm}
    \includegraphics[width=0.9\textwidth]{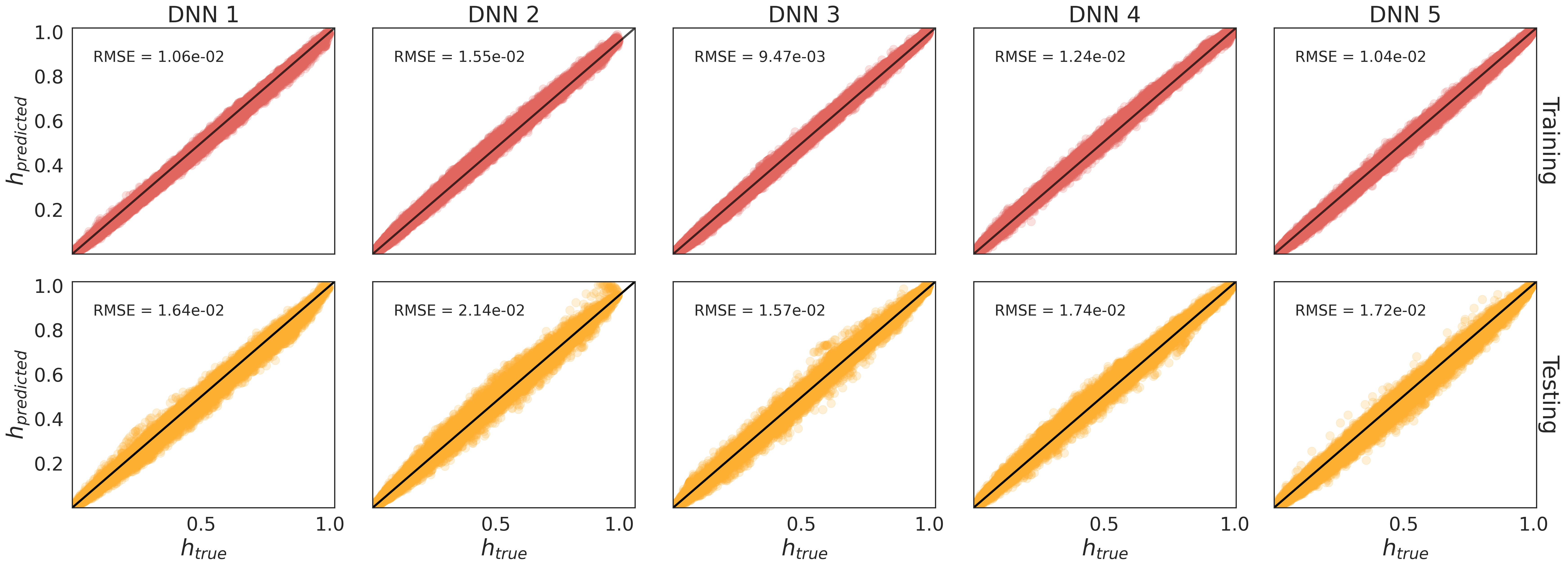}
    \caption{Performance of \ml{the five DNNs used in the multi-DNN approach, as shown in Fig. \ref{fig:multi_mlp_3d},} with respect to the training dataset (top) and the testing dataset (bottom).}
    \label{fig:multi_dnn_performance}
    \vspace{0.2cm}
\end{figure}

\subsubsection{Uncertainty Quantification}
Similarly to the first example, we chose a multivariate standard normal distribution $\pi_0(\bm{\theta}) = \mathcal{N}(0,\mathbb{I}_k)$ as the prior distribution of parameters, and set the error covariance to $\bm{\Sigma}_e = 0.001 \: \mathbb{I}_m$. \ml{Hence, the synthetic head data from the wells were pertubated with white noise with covariance $\bm{\Sigma}_e$.} In this example, we utilised the Adaptive Metropolis (AM) transition kernel for generating proposals. We completed $n=8$ independent simulations, each initialised from a random initial point close to the MAP point $\bm{\theta}_{MAP}$, with a burnin of 1000 and a final sample size of $N = 10,000$. The subchains were run with an acceptance delay of $t=2$, since longer subchains tended to diverge, leading to sub-optimal acceptance rates on the fine level. The simulations had a mean acceptance rate of 0.26, a mean effective sample size ($N_{eff}$) of 55.2 and a mean autocorrelation length $\tau = N/N_{eff}$ of 181.0. The samples of each independent simulation were pruned according to the respective autocorrelation length, and the remaining samples were pooled together to yield 443 statistically independent samples that were then analysed further.

\begin{figure}[htbp]
    \centering
    \vspace{0.2cm}
    \includegraphics[width=0.9\textwidth]{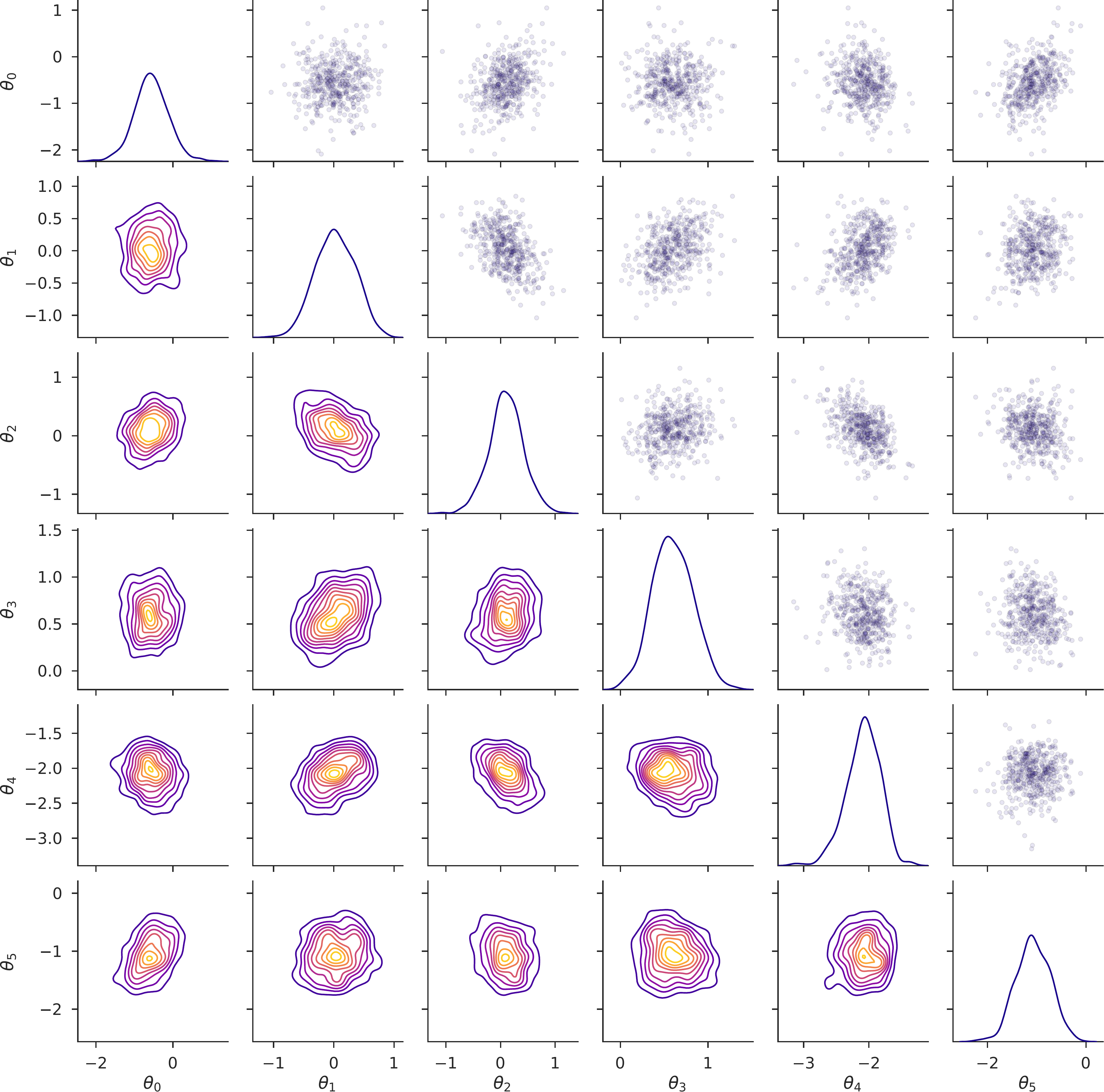}
    \caption{One and two-dimensional posterior marginal distributions (diagonal and lower triangle) and scatterplots (upper triangle) of posterior samples pruned according to the autocorrelation length of each chain for the largest 5 KL eigenmodes. Please note that the axis scales of are not equal.}
    \label{fig:parameter_matrix}
\end{figure}

Figure \ref{fig:parameter_matrix} shows the marginal distributions of the six coarsest KL coefficients along with a scatterplot matrix of all the samples remaining after pruning. All the marginal distributions are approximately Gaussian, and the two-parameter marginal distributions are mostly elliptical. It is evident that some of these parameters are correlated, namely parameters $(\bm{\theta}_0 ,\bm{\theta}_5), (\bm{\theta}_1, \bm{\theta}_2), (\bm{\theta}_1, \bm{\theta}_3),  (\bm{\theta}_1, \bm{\theta}_4)$ and $(\bm{\theta}_2, \bm{\theta}_4)$. It is worth mentioning that in every independent simulation, the AM proposal kernel managed to capture these correlations.

\begin{figure}[htbp]
    \centering
    \includegraphics[width=0.5\textwidth]{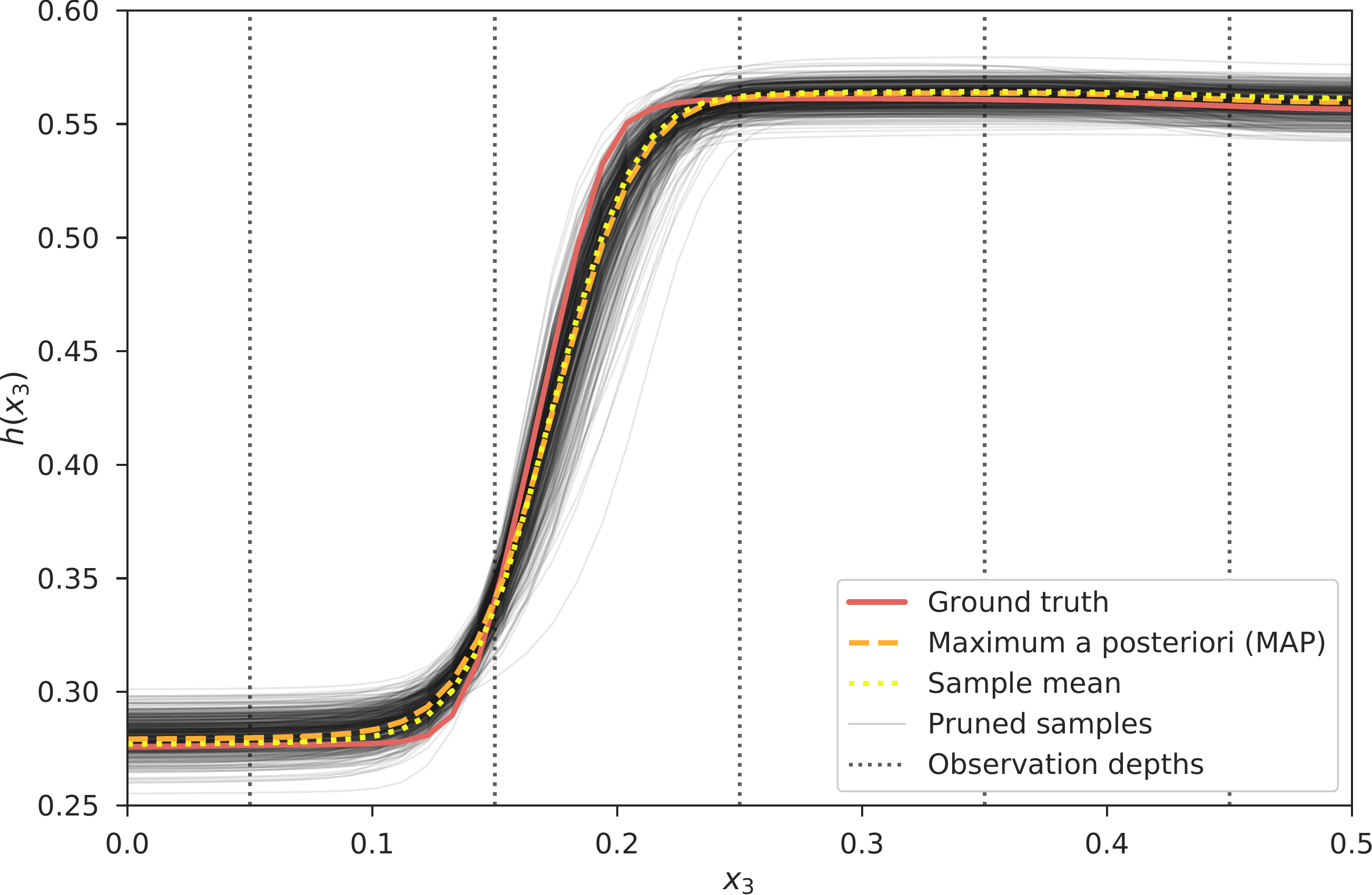}
    \caption{Hydraulic head as a function of datum $h(x_3)$ at $\bm{x} = (1.0, 0.5, x_3)^\intercal$. The solid red line shows the hydraulic head of the ground truth, the dashed orange line shows the head of the Maximum a posteriori (MAP) point $\bm{\theta}_{MAP}$, the dotted yellow line shows the mean head of the independent simulations ($n=8$) and the thin black lines show the head of 538 statistically independent samples, remaining after pruning according to the autocorrelation length of each chain, $n=443$. The vertical dotted lines show the observation depths.}
    \label{fig:head_3d}
\end{figure}

Moreover, we analysed the hydraulic head as a function of datum $h(x_3)$ along a line in the centre of the domain $\bm{x} = (1.0, 0.5, x_3)^\intercal$. Figure \ref{fig:head_3d} shows $h(x_3)$ of the ground truth, MAP point $\bm{\theta}_{MAP}$, the mean of the $n=8$ independent simulations, and all the samples remaining after pruning. We observe that both the MAP point and the sample mean are fairly close to the ground truth, albeit exhibiting higher smoothness, particularly between the observation depths, where the head is essentially allowed to vary freely. It is also clear that the individual samples encapsulate the ground truth, indicating that the ground truth is indeed contained by posterior distribution.

\section{Discussion}
\label{sec:discussion}
In this paper, we have demonstrated the use of a novel Markov Chain Monte Carlo methodology which employs a delayed acceptance (DA) model hierarchy with a deep neural network (DNN) as an approximate model and a FEM solver as a fine model, and generates proposals using the pCN and AM transition kernels. Results from the first example clearly indicate that the use of a carefully designed DNN as a model approximation can significantly reduce the cost of uncertainty quantification, even for DNNs trained on relatively small sample sizes. We have established that offsetting fine model evaluations in the DA algorithm reduces the autocorrelation of the fine chain, resulting in a higher effective sample size which, in turn, improves the statistical validity of the results. In this context, the performance of the DNN is a critical driver when determining a feasible offset length to avoid divergence of the coarse chain. Hence, if a high effective sample size is required, it may be desirable to invest in a well-performing DNN. Moreover, we have shown that an enhanced error model, which introduces an iteratively-constructed bias distribution in the coarse chain likelihood function, further increases the effective sample size and decreases the variance of the cost in repeated experiments. Finally, we observed that for the second example, even when employing a relatively well-performing model approximation, we had to constrain the offset length of the subchains rather strongly to achieve optimal acceptance rates. This can be attributed in part to an apparent non-spherical and correlated posterior distribution, causing the employed proposal kernels to struggle to discover areas of high posterior probability.

\smallskip
We have demonstrated that relatively simple inverse hydrogeological problems can be solved in reasonable time on a commonly available personal computer with no GPU-acceleration. This opens the opportunity to apply robust uncertainty quantification during fieldwork and as a decision-support tool for groundwater surveying campaigns. We have also demonstrated the applicability of our approach on a larger scale three-dimensional problem, utilising a GPU-accelerated high-performance computer (HPC). Aside from the benefit of using a HPC computer for accelerating the fine model evaluations, utilising the GPU allowed for rapidly training and testing multiple different DNN designs to efficiently establish a well performing model approximation. There are other obvious ways to further increase the efficiency of the proposed methodology. For example, construction of the DNNs used as coarse models comes with the cost of evaluating multiple models from the prior distribution, and, unlike the MCMC sampler, the prior models are independent and these fine model evaluations can thus be massively parallelised.

\smallskip
Our methodology was demonstrated in the context of two relatively simple groundwater flow problems with log-Gaussian transmissivity fields parametrised by Karhunen-Loève decompositions. While this model provides a convenient computational structure for our purposes, it may not reflect the full scale transmissivity of real-world aquifers, particularly in the presence of geological faults and other heterogeneities, as discussed in \cite{gomez-hernandez_be_1998}. Future research could address this problem through geological layer stratification using the universal cokriging interpolation method suggested in \cite{lajaunie_foliation_1997}, potentially utilising the open-source geological modelling tool GemPy \cite{de_la_varga_gempy_2019}, which allows for simple parametric representation of geological strata. Spatially heterogeneous parameters within each strata could then be modelled hierarchically using a low order log-Gaussian random field to account for within-stratum variation, as demonstrated in \cite{mondal_bayesian_2010}.

\section{Acknowledgements}
This work was funded as part of the Water Informatics Science and Engineering Centre for Doctoral Training (WISE CDT) under a grant from the Engineering and Physical Sciences Research Council (EPSRC), grant number EP/L016214/1. TD was funded by a Turing AI Fellowship (2TAFFP\textbackslash100007). DM acknowledges support from the EPSRC Platform Grant PRISM (EP/R029423/1). The authors have no competing interests. Data supporting the findings in this study are under preparation and will be made available in the Open Research Exeter (ORE, \textit{https://ore.exeter.ac.uk/repository/}) data repository.

\bibliographystyle{elsarticle-num}  
\bibliography{main}

\appendix

\newpage
\section{Preconditioned Crank-Nicolson} \label{app:pCN}
The preconditioned Crank-Nicolson (pCN) proposal was developed in \cite{cotter_mcmc_2013} and is based on the following Stochasic Partial Differential Equation (SPDE):

\begin{equation*}
\frac{du}{ds} = - \mathcal{K} \mathcal{L} u + \sqrt{2 \mathcal{K}} \frac{db}{ds}
\end{equation*}

where $\mathcal{L} = \mathcal{C}^{-1}$ is the precision operator for the prior distribution $\mu_0$, $b$ is brownian motion with covariance operator $I$, and $\mathcal{K}$ is a positive operator. This equation can be discretised using the Crank Nicolson approach to yield

\begin{equation*}
v = u - \frac{1}{2} \delta \mathcal{K} \mathcal{L} (u+v) + \sqrt{2 \mathcal{K} \delta} \xi_0
\end{equation*}

for white noise $\xi_0$ and a weight $\delta \in [0,2]$. If we choose $\mathcal{K} = I$, we get the plain Crank Nicolson (CN) proposal:

\begin{equation*}
(2 \mathcal{C} + \delta I) v = (2\mathcal{C} - \delta I) u + \sqrt{8 \delta \mathcal{C}} \xi
\end{equation*}

where $\xi \sim \mathcal{N}(0,\mathcal{C})$. If we instead choose $\mathcal{K} = \mathcal{C}$, we get the pCN proposal:

\begin{equation*}
v = \sqrt{1 - \beta^2} u + \beta \xi, \quad
\beta = \frac{\sqrt{8 \delta}}{2 + \delta}, \quad \beta \in [0,1]
\end{equation*}

This is rewritten, conforming to our previous notation:

\begin{equation*}
\bm{\theta}' = \sqrt{1-\beta^2}\bm{\theta}_i + \beta \bm{\xi}
\end{equation*}

\newpage
\section{\ml{Recovered Conductivity Fields}} \label{app:recovered}

\begin{figure}[htbp]
    \centering
    \vspace{0.2cm}
    \includegraphics[width=0.8\textwidth]{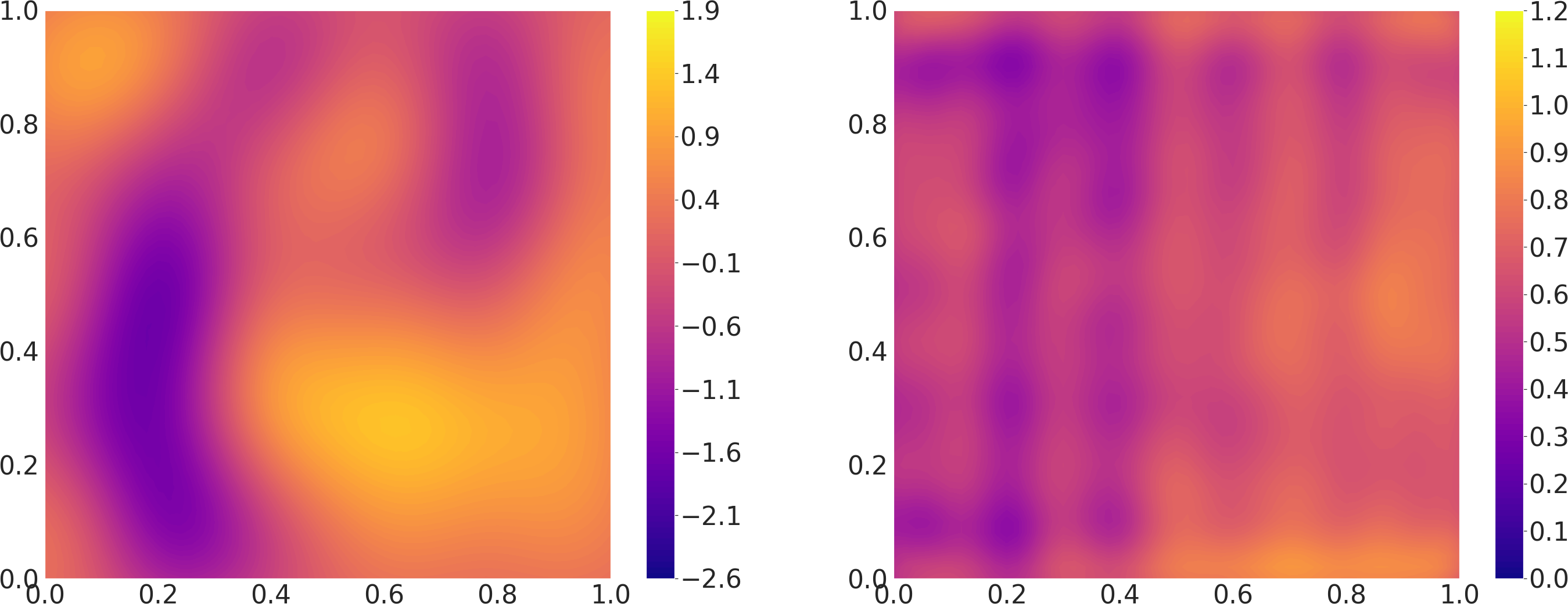}
    \caption{Mean (left) and variance (right) of recovered log-transmissivity for Vanilla pCN.}
    \label{fig:recovered_mean__and_variance_vanilla_pcn}
\end{figure}

\begin{figure}[htbp]
    \centering
    \vspace{0.2cm}
    \includegraphics[width=0.8\textwidth]{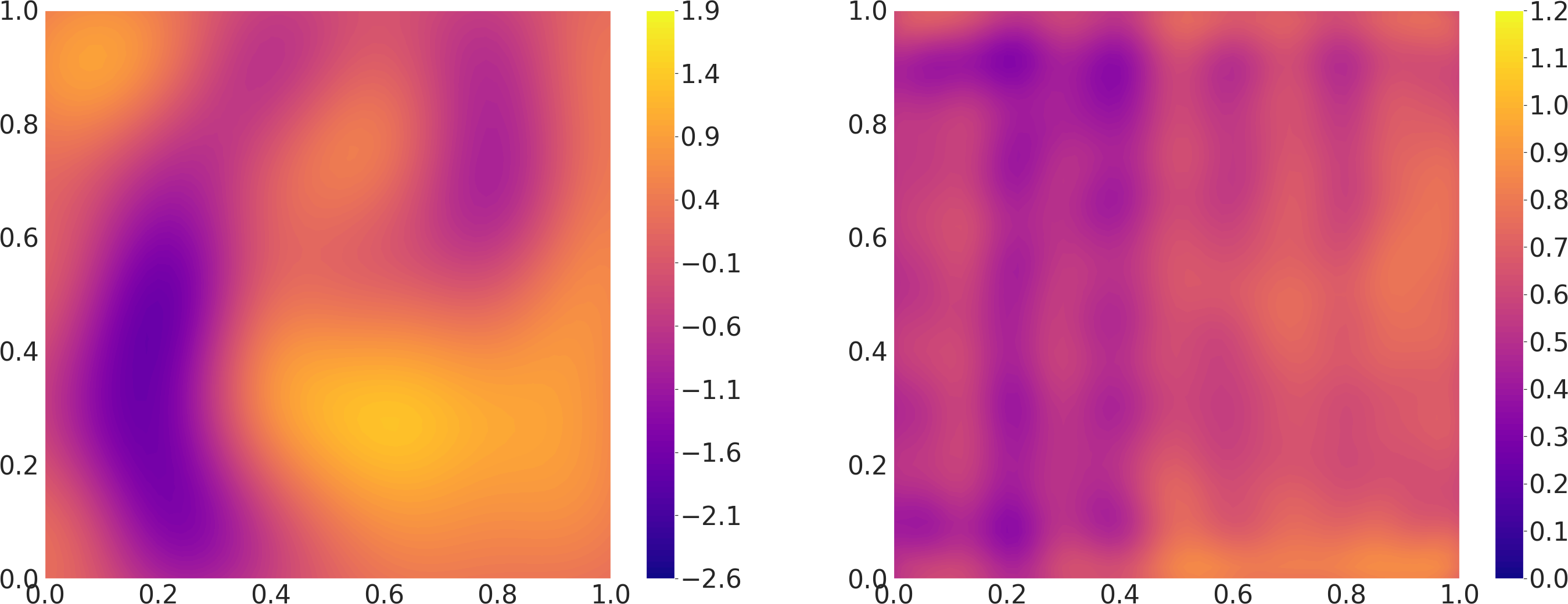}
    \caption{Mean (left) and variance (right) of recovered log-transmissivity for DA, $N_{DNN} = 4000$.}
    \label{fig:recovered_mean__and_variance_dnn_4000}
\end{figure}

\begin{figure}[htbp]
    \centering
    \vspace{0.2cm}
    \includegraphics[width=0.8\textwidth]{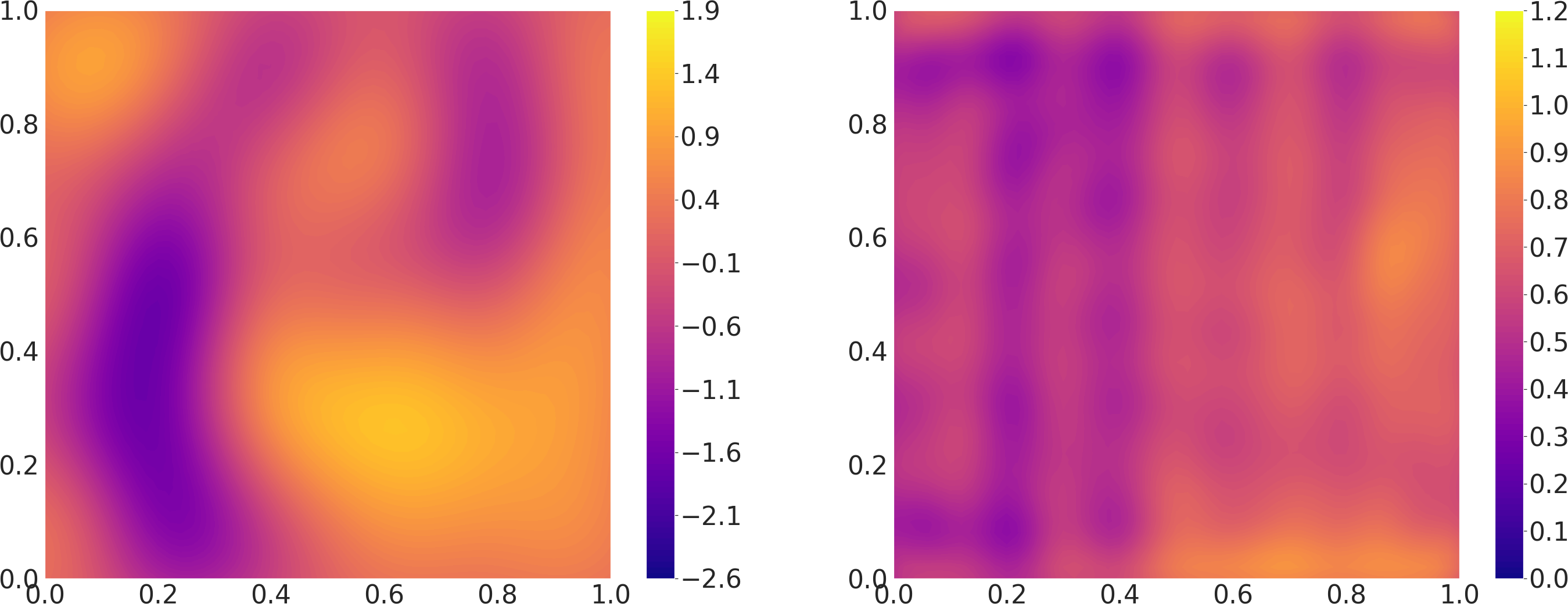}
    \caption{Mean (left) and variance (right) of recovered log-transmissivity for DA/EEM, $N_{DNN} = 4000$.}
    \label{fig:recovered_mean__and_variance_dnn_4000_eem}
\end{figure}

\begin{figure}[htbp]
    \centering
    \vspace{0.2cm}
    \includegraphics[width=0.8\textwidth]{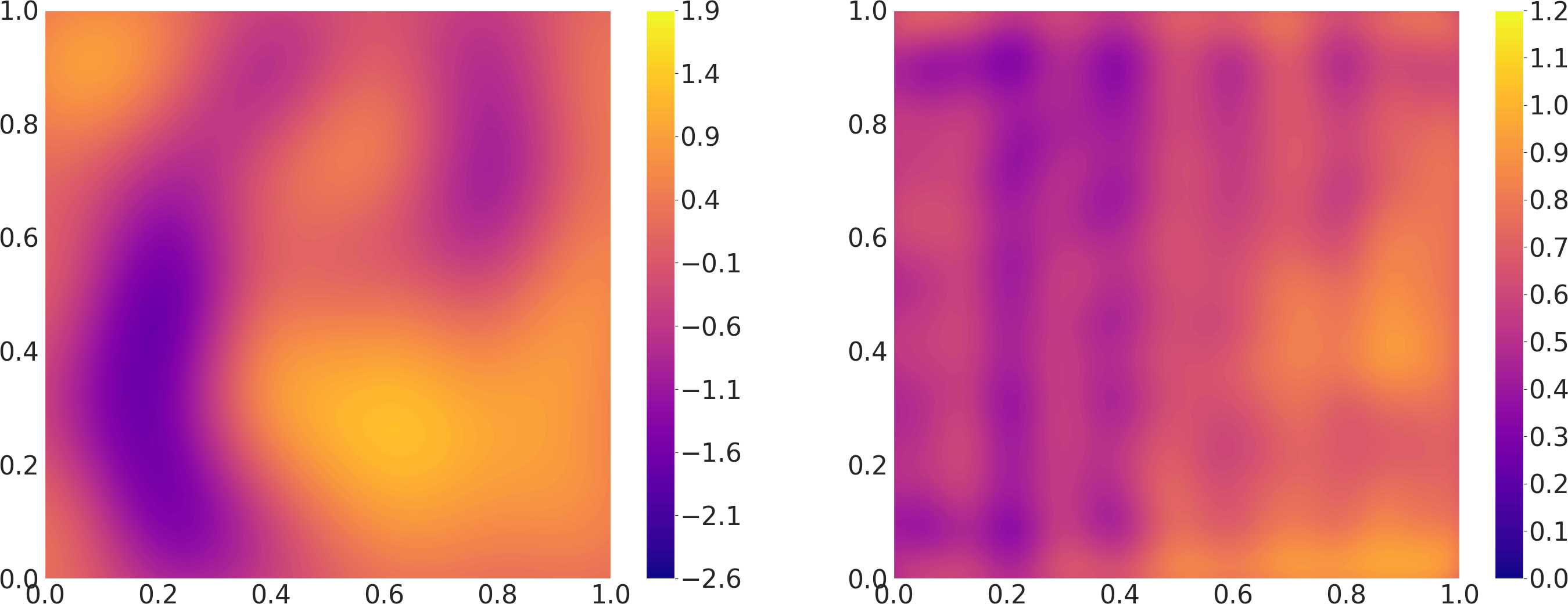}
    \caption{Mean (left) and variance (right) of recovered log-transmissivity for DA, $N_{DNN} = 16000$.}
    \label{fig:recovered_mean__and_variance_dnn_16000}
\end{figure}

\begin{figure}[htbp]
    \centering
    \vspace{0.2cm}
    \includegraphics[width=0.8\textwidth]{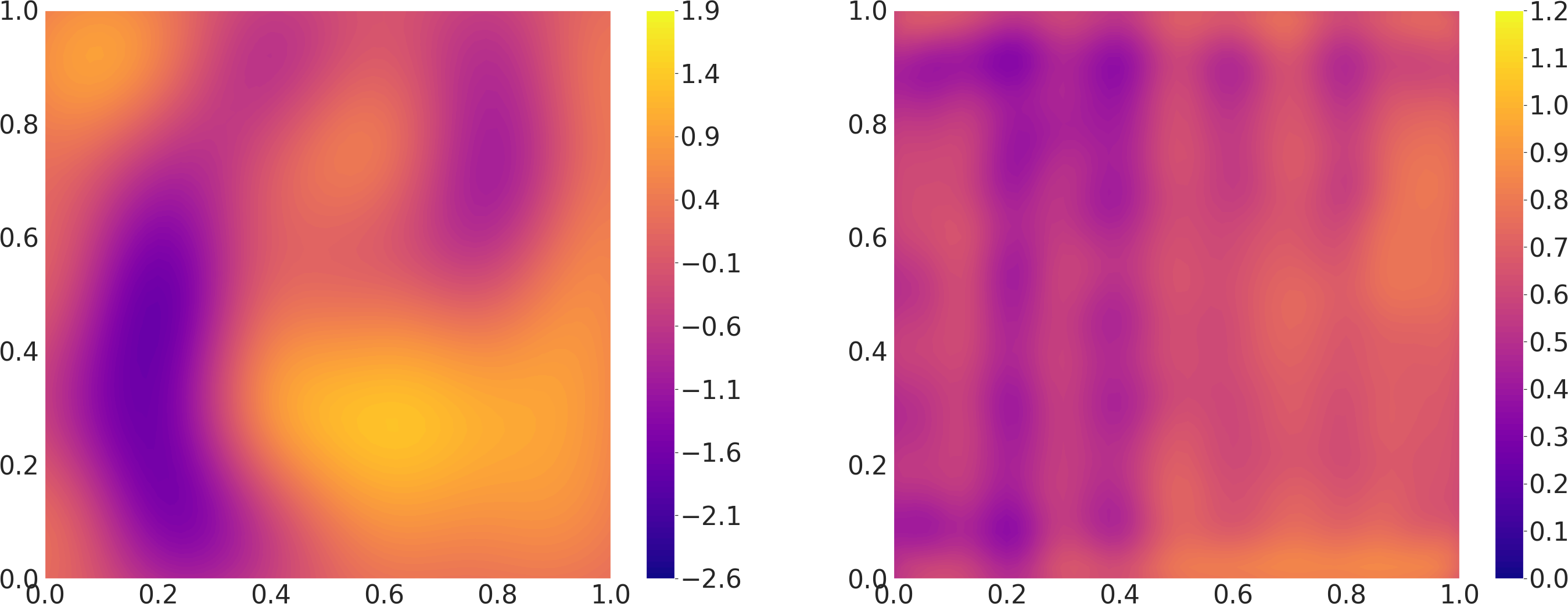}
    \caption{Mean (left) and variance (right) of recovered log-transmissivity for DA/EEM, $N_{DNN} = 16000$.}
    \label{fig:recovered_mean__and_variance_dnn_16000_eem}
\end{figure}

\begin{figure}[htbp]
    \centering
    \vspace{0.2cm}
    \includegraphics[width=0.8\textwidth]{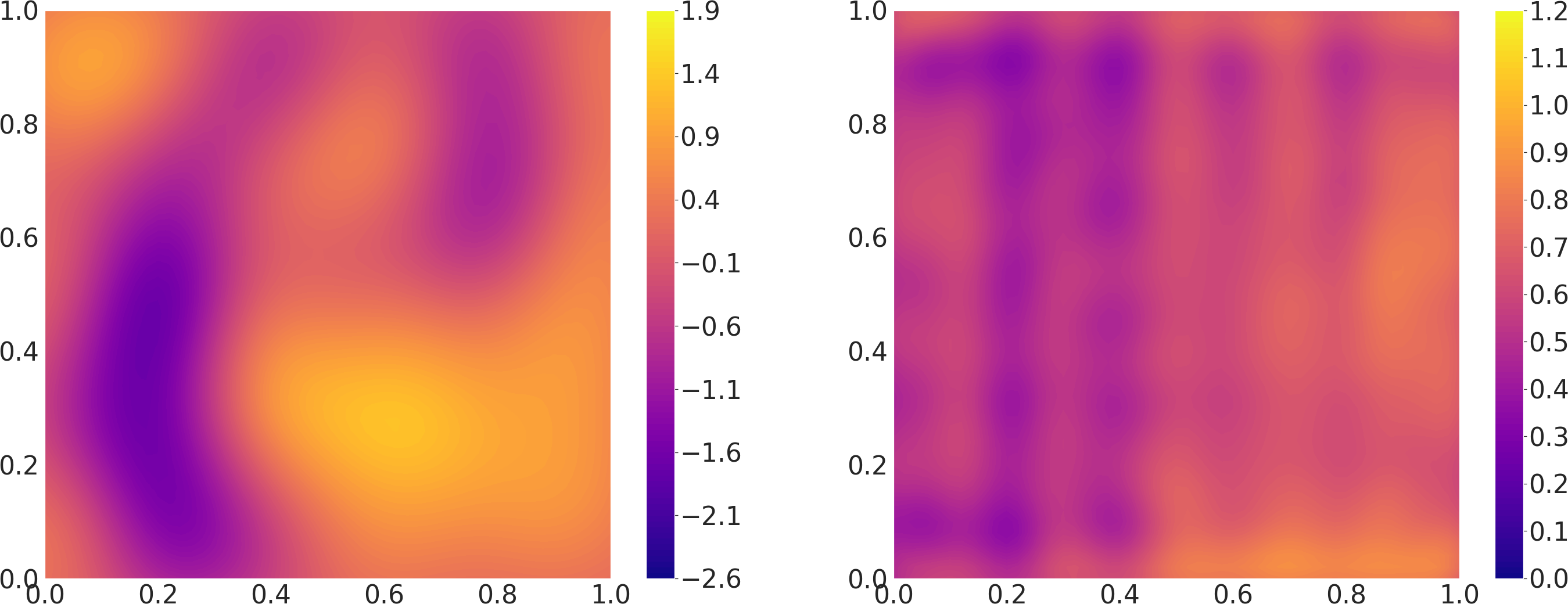}
    \caption{Mean (left) and variance (right) of recovered log-transmissivity for DA, $N_{DNN} = 64000$.}
    \label{fig:recovered_mean__and_variance_dnn_64000}
\end{figure}

\begin{figure}[htbp]
    \centering
    \vspace{0.2cm}
    \includegraphics[width=0.8\textwidth]{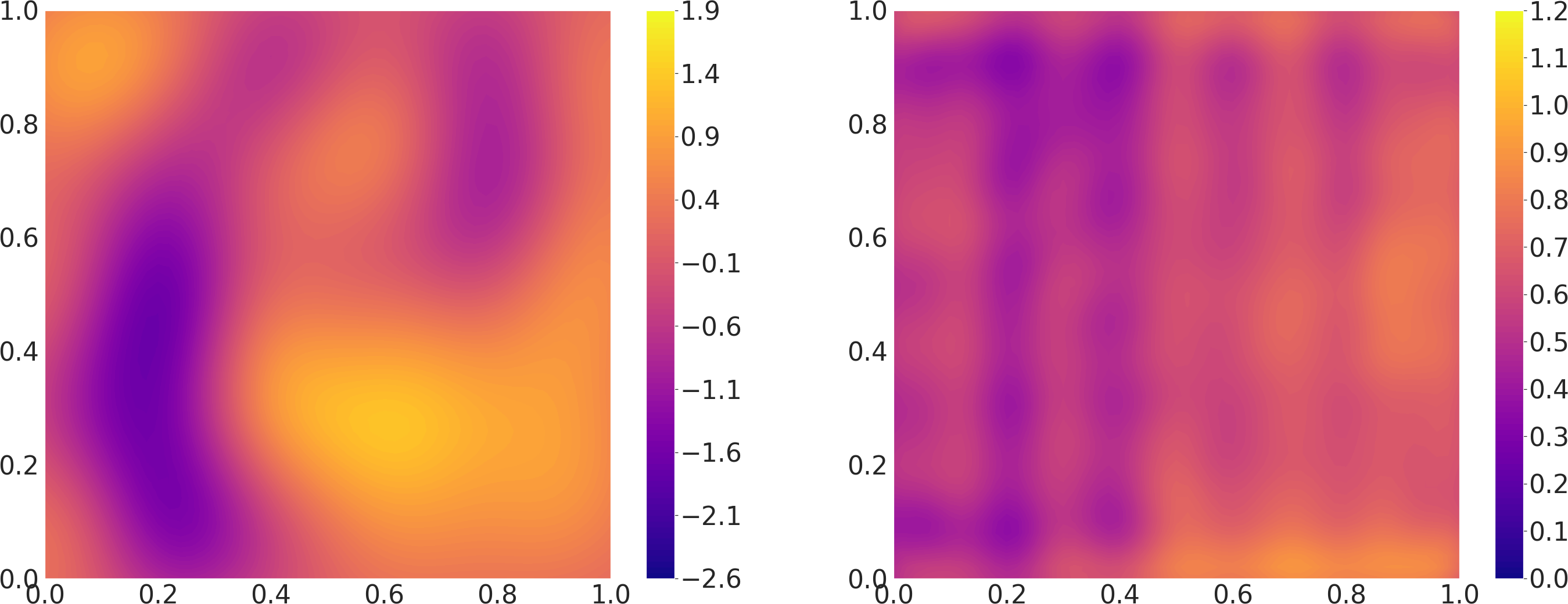}
    \caption{Mean (left) and variance (right) of recovered log-transmissivity for DA/EEM, $N_{DNN} = 64000$.}
    \label{fig:recovered_mean__and_variance_dnn_64000_eem}
\end{figure}

\end{document}